\documentclass[]{aastex631}

\usepackage{amsmath}
\usepackage{hyperref}
\begin{document}

\title{Casting Light on Degeneracies: A Comprehensive Study of Lightcurve Variations in Microlensing Events OGLE-2017-BLG-0103 and OGLE-2017-BLG-0192}

\correspondingauthor{Sarang Shah}

\email{sshah1502@gmail.com}

\author[0000-0003-1959-8439]{Sarang Shah}

\affil{Indian Institute of Astrophysics, Koramangala, Bengaluru, India - 560034}

\begin{abstract}
    \noindent This study investigates orbital parallax in gravitational microlensing events, focusing on OGLE-2017-BLG-0103 and OGLE-2017-BLG-0192. For events with timescales $\leq$ 60 days, a Jerk Parallax degeneracy arises due to high Jerk velocity ($\tilde{v_{j}}$), causing a four-fold continuous parallax degeneracy. OGLE-2017-BLG-0103, after incorporating orbital parallax, reveals four discrete degenerate parallax solutions, while OGLE-2017-BLG-0192 exhibits four discrete solutions without degeneracy. {The asymmetric light curve of OGLE-2017-BLG-0103 suggests a more probable model where xallarap is added to the parallax model, introducing tension. The galactic model analysis predicts a very low mass stellar lens for OGLE-2017-BLG-0192. For OGLE-2017-BLG-0103, degenerate solutions suggest a low-mass star or a darker lens in the disc, while the Xallarap+Parallax model also predicts a stellar lens in the bulge, with the source being a solar-type star orbited by a dwarf star.} This study presents five degenerate solutions for OGLE-2017-BLG-0103, emphasizing the potential for confirmation through high-resolution Adaptive Optics (AO) observations with Extremely Large Telescopes in the future. The complexities of degenerate scenarios in these microlensing events underscore the need to analyze special single-lens events in the Roman Telescope Era.
\end{abstract}\vspace{-2em}
\section{Introduction}
\begin{figure}
\centering
\includegraphics[width=0.45\textwidth]{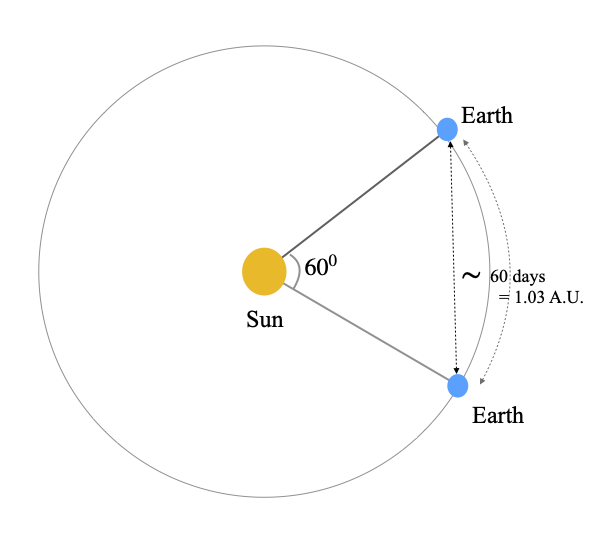}
  \caption{The projected $r_{E}$ for the lenses towards the bulge on the ecliptic plane is $\sim$ 1.03 A.U. Earth also displaces by this value in 60 days. The displacement equal to the projected Einstein radius of the lens helps to measure the orbital parallax.}
  \label{fig:1.9}
 \end{figure}

\noindent Gravitational microlensing \citep{einstein1936, paczynski1986} is an intriguing astronomical phenomenon where the gravity of a foreground lens causes the light from a background source to bend, resulting in the appearance of multiple images when their relative proximity in the sky is closer than the $\theta_{E}$ which is given by,
\begin{equation}
    \theta_{E} = \sqrt{\frac{4GM_{L}}{c^{2}}\Big(\frac{1}{D_{L}}- \frac{1}{D_{S}}}\Big)
    \label{eq:t0}
\end{equation}
where 4G/$c^{2}$ $\sim$ 8.14 $mas$$M_{\odot}^{-1}$, $M_{L}$ is the mass of the lens, and $D_{L}$ and $D_{S}$ are the distances to the lens and source, respectively. These images are not individually resolved, distinguishing microlensing from $Gravitational$ $Lensing$. Instead, microlensing is detectable as a temporary fluctuation in the brightness of the source star when continuously observed over days or months. { This phenomenon produces a characteristic light curve known as the {\it Paczy\`nski Curve} \citep{paczynski1986} when plotted over time}. It was only through the establishment of dedicated survey groups such as OGLE-I \citep{udalski1994}, EROS \citep{aubourg1994}, and MACHO \citep{bennett1993}, that the first microlensing event toward the galactic bulge was observed \citep{alcock1995, alcock1995a}. Since then, this field has rapidly expanded, shedding light on the mysteries of the inner Milky Way, thanks to the latest generation of survey telescopes like OGLE-IV \citep{udalski2015b}, KMTNet \citep{kim2016} and MOA-III \citep{bond2001, sumi2003} equipped with a wide-field camera.\\
 A straightforward way to parameterize a microlensing event is to consider a point lens, a point source, and the observer-lens-source combination moving in relative motion. In such a scenario, the separation between the lens and the source, normalized to the angular Einstein radius of the lens, is given by:
\begin{equation}
    u(t) = \sqrt{u^{2}_{0} + \left(\frac{t - t_{0}}{t_{E}}\right)^{2}}
    \label{eq:a0}
\end{equation}
\noindent Here, $u(t)$ is a function of time $t$, $u_{0}$ and $t_{0}$ represent the impact parameters and the time of the closest approach, and $t_{E}$ corresponds to the time the source takes to cross the Einstein Radius. The time-varying magnification is then given by:
\begin{equation}
    A(t) = \frac{u^{2} + 2}{u\sqrt{u^{2} + 4}}
    \label{eq:a1}
\end{equation}
\noindent Although the lens itself cannot be directly detected during a microlensing event, higher-order effects can sometimes be observed, leading to the characterization of the lens. One such effect is known as {\it Microlensing Parallax} ($\mathbf{\pi_{E}}$), which is a vector quantity having a magnitude equal to $\tilde{r}_{E}$ and a direction similar to the project lens-source proper motion ($\mu_{rel}$) (Equation (\ref{eq:1.19})). { {Orbital parallax causes the source to follow a non-rectilinear trajectory, resulting in an asymmetric light curve, which becomes significant for events lasting 60 days or longer} \citep{gould1994, alcock1995, mao2002, soszynski2001, bond2002, bennett2002a, bennett2002b, smith2003b}. During this period, the Earth's displacement around the Sun is approximately 1.03 A.U. (see Figure (\ref{fig:1.9})). The typical value of $\tilde{r}_{E}$ for M-dwarf lenses towards the bulge is around 2 A.U., which, while larger, is comparable to the chord length depicted in Figure \ref{fig:1.9}. For events with shorter timescales ($t_{E}$), the chord length is much smaller than 1.03 A.U., rendering $\pi_{E}$ unconstrained due to negligible perturbations in the light curve. However, for longer-duration events where the displacement exceeds the chord length, $\pi_{E}$ can be effectively constrained, resulting in a noticeably asymmetric light curve. Alternatively, even if the timescale of an event is larger than 60 days, but the lens has a low mass, then the value of $\tilde{r}_{E}$ will be smaller, and hence the $\pi_{E}$ won't be constrained properly eventually leading to degenerate solutions.}\\
\begin{equation}
    \mathbf{\pi_{E}} = \frac{1}{{\mathbf{\tilde{r}_{E}}}} \, \text{A.U.}
    \label{eq:1.19}
\end{equation}
\noindent Note that the $\pi_{E}$ is a vector quantity and the direction of the $\pi_{E}$ vector is similar to the direction of the lens-source proper motion ($\mathbf{\mu_{rel}}$ = $\mathbf{\theta_{E}}$/$t_{E}$) vector \citep{gould1992b}. { Then, the mass of the lens can be calculated using,}
\begin{equation}
M_{L} = \frac{\pi_{rel}}{\kappa{\pi_{E}^{2}}} = \frac{\theta_{E}}{\kappa{\pi_{E}}}
\label{eq:1.20}
\end{equation}
where $\pi_{rel} = \text{A.U.}\left(\frac{1}{D_{L}}-\frac{1}{D_{s}}\right)$ is the relative lens-source parallax. So, the mass of the lens can be measured if we know the value of $\pi_{rel}$ or $\theta_{E}$ in addition to $\pi_{E}$. Since the lens is usually faint and distant, we cannot measure $D_{L}$ (and hence $\pi_{rel}$) directly. The parameter $\theta_{E}$ is also obtained from the lightcurve when finite-source effects are measured (see \cite{gould1994v, witt1995,nemiroff1994}). Interestingly, only a handful of single-lens events have 
directly measured the value of $\theta_{E}$ either by measuring finite-source effects in extremely high magnification events \citep{yoo2004, gould2009}, via resolution of microlensing images using interferometry \citep{delplancke2001,dong2019}, or via astrometric-microlensing \citep{hog1995, sahu2022}. As these measurements are not possible for every microlensing event, measuring $\pi_{E}$ signatures in the lightcurve helps to put constraints on $M_{L}$ and $D_{L}$.\\
For microlensing events having timescales $\sim$ 60 days, \citet{gould2004} showed that they suffer from a four-fold continuous degeneracy called as {\it Jerk Parallax} (${\pi_{j}}$) degeneracy. This degeneracy is important when the value of $\tilde{v_{j}}$ is high and $t_{E}$ is $\sim$ 60 days. As $t_{E}$ increases, the continuous degeneracy is broken into discrete degeneracy and completely broken for very long $t_{E}$ events. We encourage the readers to refer \cite{smith2003a, gould2004} to understand the origins and fundamentals of this degeneracy. While this degeneracy has been explored before by \cite{gould2004, jiang2004, park2004, Poindexter2005}, { exactly, how does this degeneracy break as $t_{E}$ increases, and how do multi-site observations help in realizing this degeneracy has never been seen before. While the events analyzed by the first three authors had timescales $\leq$ 60 days, the latter found degenerate solutions in events having $t_{E}$ higher than 60 days.}\\
\noindent In this paper, we present our analysis of the microlensing events OGLE-2017-BLG-0103 and OGLE-2017-BLG-0192. By modeling the light curve involving $\pi_{E}$, we identify the existence of a discrete Jerk-Parallax degeneracy in the light curve of the former event but not later. We also explore the possibility of the source having an orbiting companion causing effects (also called xallarap) similar to the observed orbital parallax effect. For OGLE-2017-BLG-0103, we find a strong degeneracy between the orbital parallax and the Xallarap+Parallax model. We do not find such tension in OGLE-2017-BLG-0192.
\subsection{Overview}\label{section:overview}
\noindent After an introduction, we discuss the details of the data used in this analysis (\ref{section:datareductions}). { In the initial part of Section (\ref{section:analysis}), we describe the microlensing model that we fit the observed data followed by sub-Sections (\ref{section:op_model}) and (\ref{section:alternate}), where we talk about our approach to finding the Jerk-Parallax degenerate solutions (both giving similar results), followed by Section (\ref{section:xallarap}) where we discuss our Xallarap model.} In Section (\ref{section:cmd}), we analyze the source and the blend, followed by Section (\ref{seection:lens}), where we obtain the distance to and mass of the lens. Finally, we discuss our results and conclude in Section (\ref{section:discussion}).
\section{Observations and Data Reductions}\label{section:datareductions}
\noindent OGLE-2017-BLG-0103 is the 103\textsuperscript{rd} event detected by the Early Warning System (EWS) of OGLE in the field \textit{BLG501.11} while OGLE-2017-BLG-0192 is the 192\textsuperscript{nd} detected in the OGLE field \textit{BLG518.09}. It was observed towards the galactic bulge through its 1.3m Warsaw Telescope, located at the Las Campanas Observatory in Chile \cite{udalski2015b}. OGLE has a wide-field CCD camera, a mosaic of 32 2k x 4k individual chips. Together, they contribute to the camera having a wide FoV of 1.4 sq. deg. { The calibrated source in both these events is a faint star with a baseline magnitude of $I$ $\geq$ 18.5 mag on the OGLE photometric maps. The equatorial co-ordinates of OGLE-2017-BLG-0103 are $\alpha = 17:52:31.49$, $\delta = -30:00:44.4$ which can be translated to galactic co-ordinates as $l=359.85008^{o}$, $b = -1.84077^{o}$. Similarly, the equatorial coordinates for OGLE-2017-BLG-0192 are $\alpha = 18:10:34.45$, $\delta = -26:50:25.9$, which can be translated to $(l,b)$ = ($4.57154^{o}, -3.7277^{o}$). We download the OGLE photometry for both events, which is available on its homepage. This photometry is generated after performing the difference image analysis of the field images that use the algorithm by \cite{alard1998}. Both these microlensing events were also observed independently by KMTNet through its three 1.8m telescopes spread across longitude in the southern hemisphere in Chile (KMT-C), South Africa (KMT-S), and Australia (KMT-A). {  Both these events were also found during the 2016 season by KMTNet's event finder algorithm \citep{kim2018} and were designated in the KMTNet database as KMT-2017-BLG-1698 and KMT-2017-BLG-1225.} All three KMTNet telescopes have $4 deg^{2}$ wide-field cameras, each with four 9k x 9k chips to monitor the galactic field \cite{kim2016}. While OGLE-2017-BLG-0103 was observed in two fields - BLG02 and BLG42, OGLE-2017-BLG-0192 was observed in the field BLG31 by the KMTNet. Most of the observations were taken in the I band; there are occasional V band observations for these events. The KMTNet images in the I and V bands were reduced using the pyDIA software \cite{albrow2017}, based on difference image analysis using the delta-basis-function approach of \cite{bramich2013}.}
\section{Analysis}\label{section:analysis}
\noindent As mentioned in the introduction, a single lens and single source lightcurve is described by three Pa\`czynski parameters: $u_{0}$, $t_{0}$, and $t_{E}$. Two additional fit parameters, the $f_{s}$ and $f_{b}$, are the source and blend fluxes obtained by performing a linear fit of the model to the observed data (refer Equation (\ref{eq:n1})). However, as we describe later, we incorporate a few additional parameters and form a higher-order model to fit the asymmetry in the data. We model the observed light curve of both events using the I band data of OGLE and KMTNet. So we have fourteen and eight values of $f_{s}$ and $f_{b}$ for OGLE-2017-BLG-0103 and OGLE-2017-BLG-0192, respectively, in addition to three Paczy\`nski parameters to fit the observed light curve.
\begin{equation}
    Model(t) = A(t)f_{s} + f_{b}
    \label{eq:n1}
\end{equation}
\subsection{Orbital Parallax Model}\label{section:op_model}
\begin{figure*}
\centering
\includegraphics[width=0.45\textwidth]{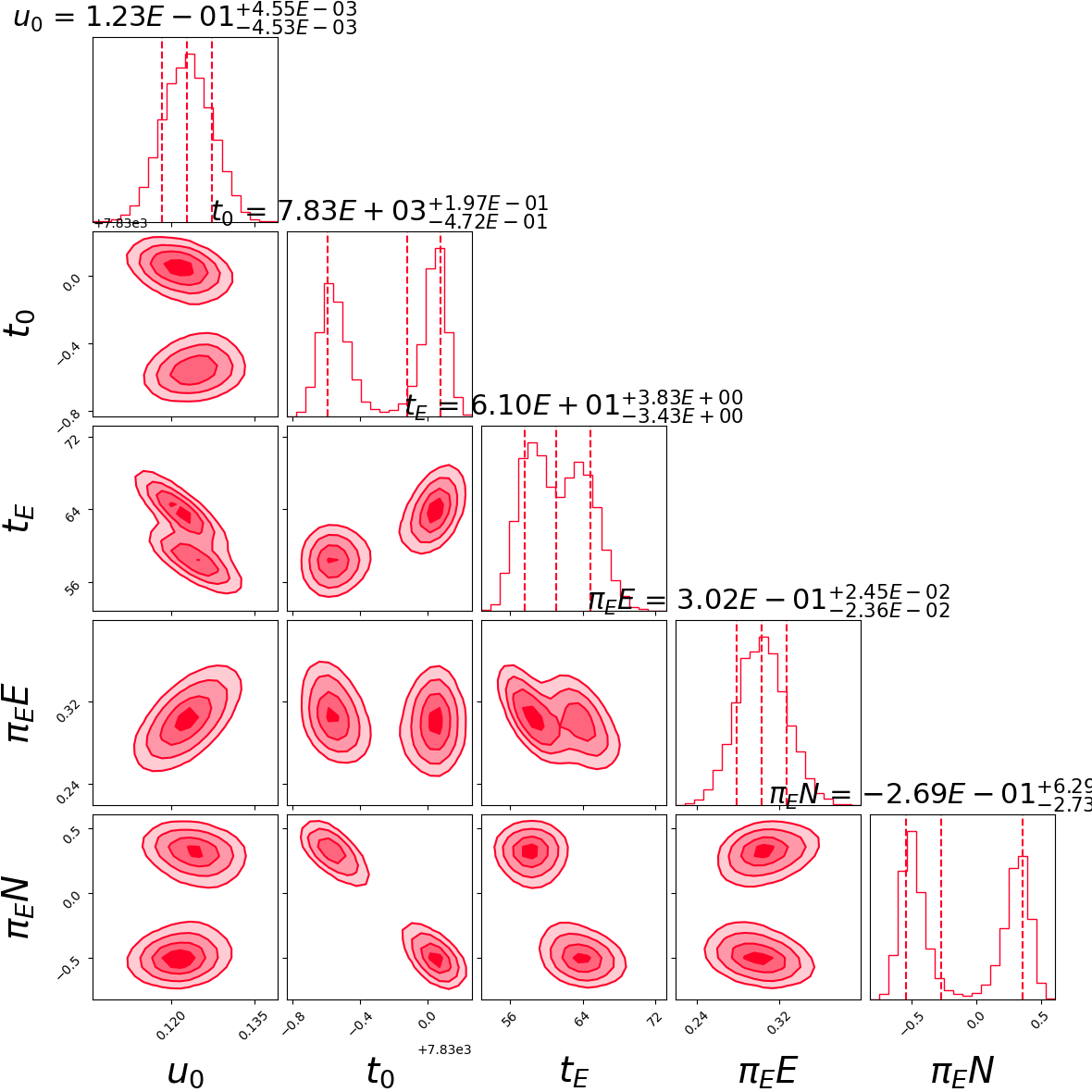}
\includegraphics[width=0.45\textwidth]{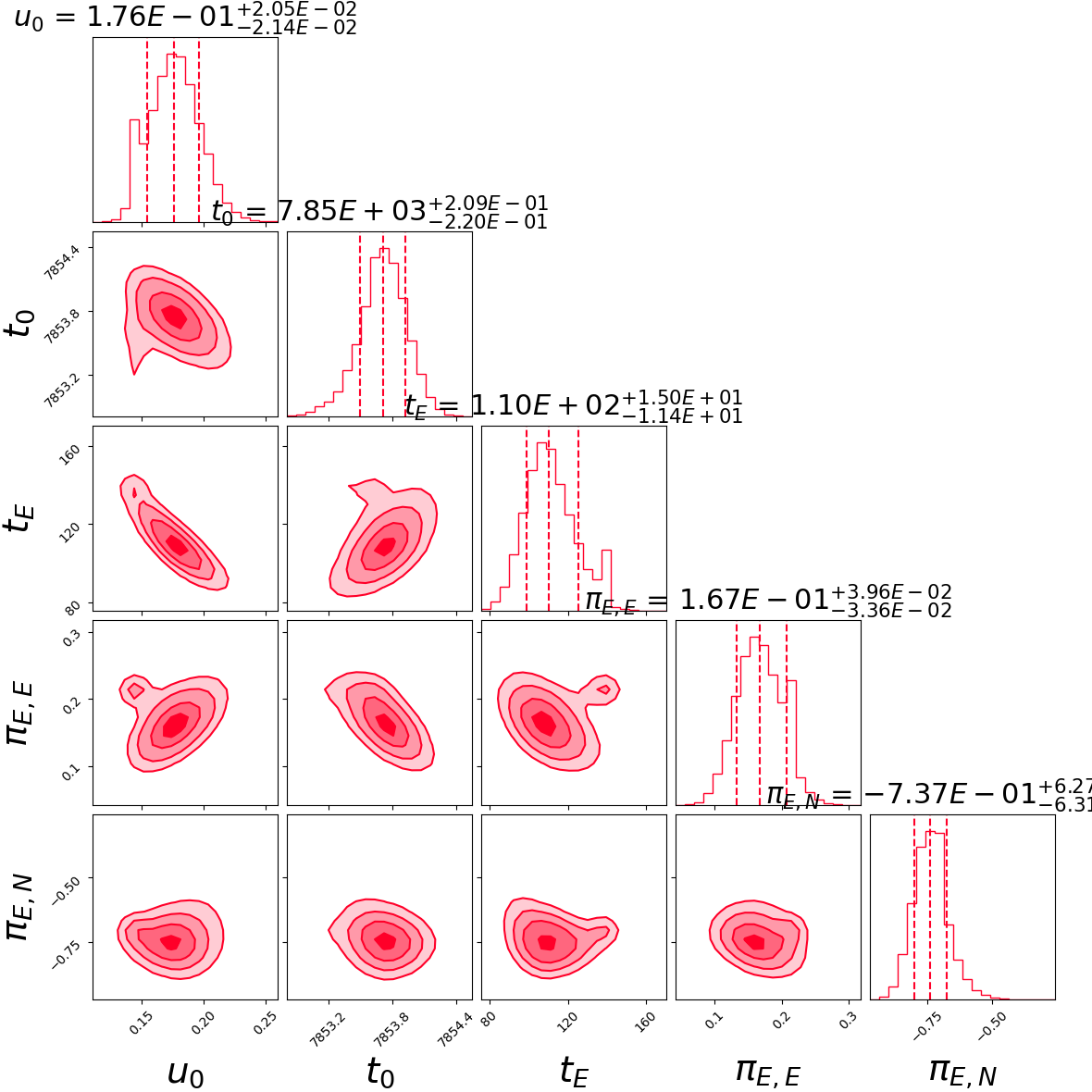}
\caption{The covariance plots of the model parameters { $\mathbf{left}$} OGLE-2017-BLG-0103 and {$\mathbf{right}$} OGLE-2017-BLG-0192. In the case of OGLE-2017-BLG-0103, we can see the Jerk-Parallax degeneracy in the bimodal distributions of $\pi_{EN}$, $t_{0}$ and $t_{E}$.}
\label{fig:triangle_both}
\end{figure*}

\begin{figure*}[!htb]
  \centering
\includegraphics[width=0.45\textwidth]{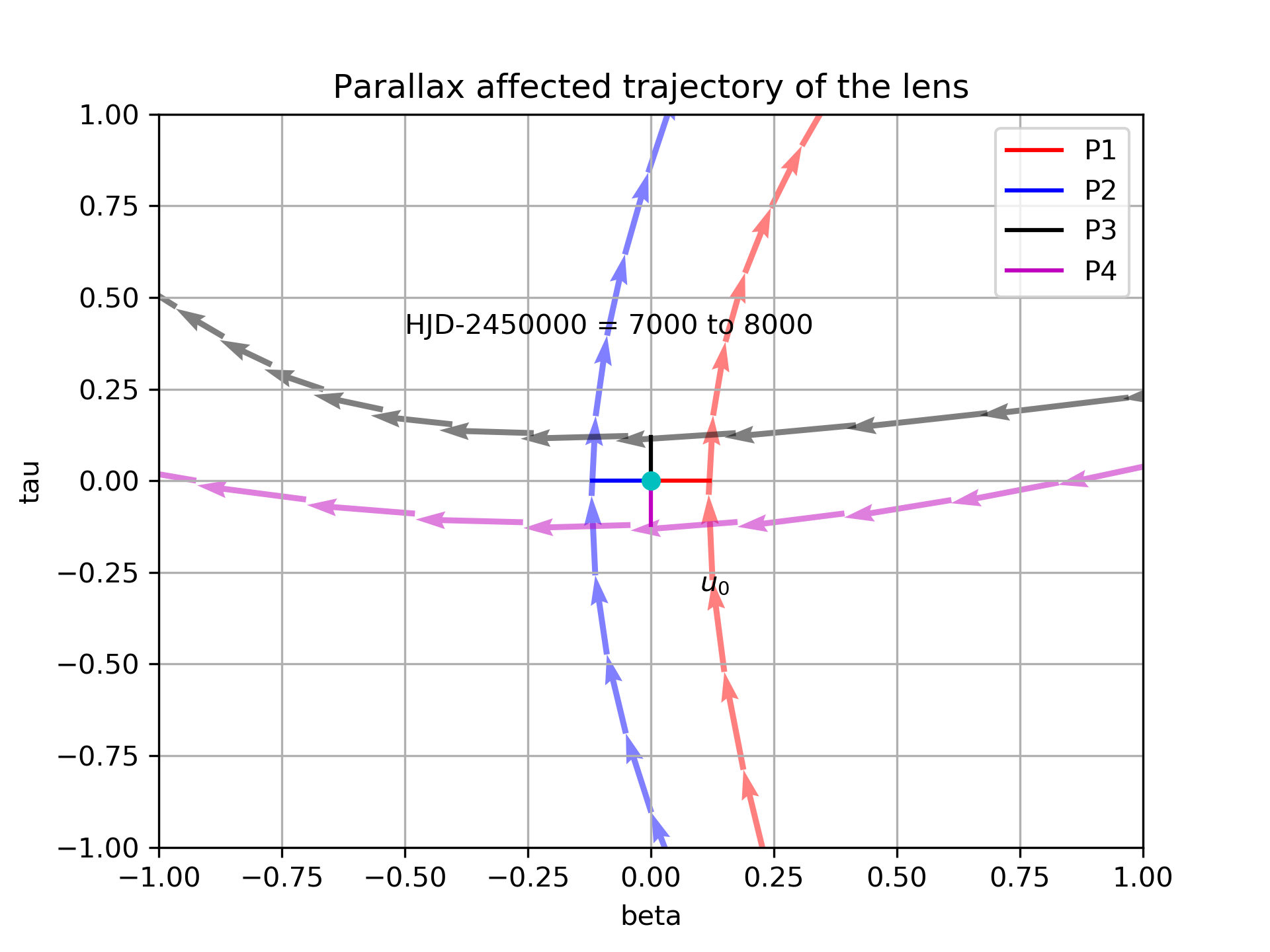}
\includegraphics[width=0.50\textwidth]{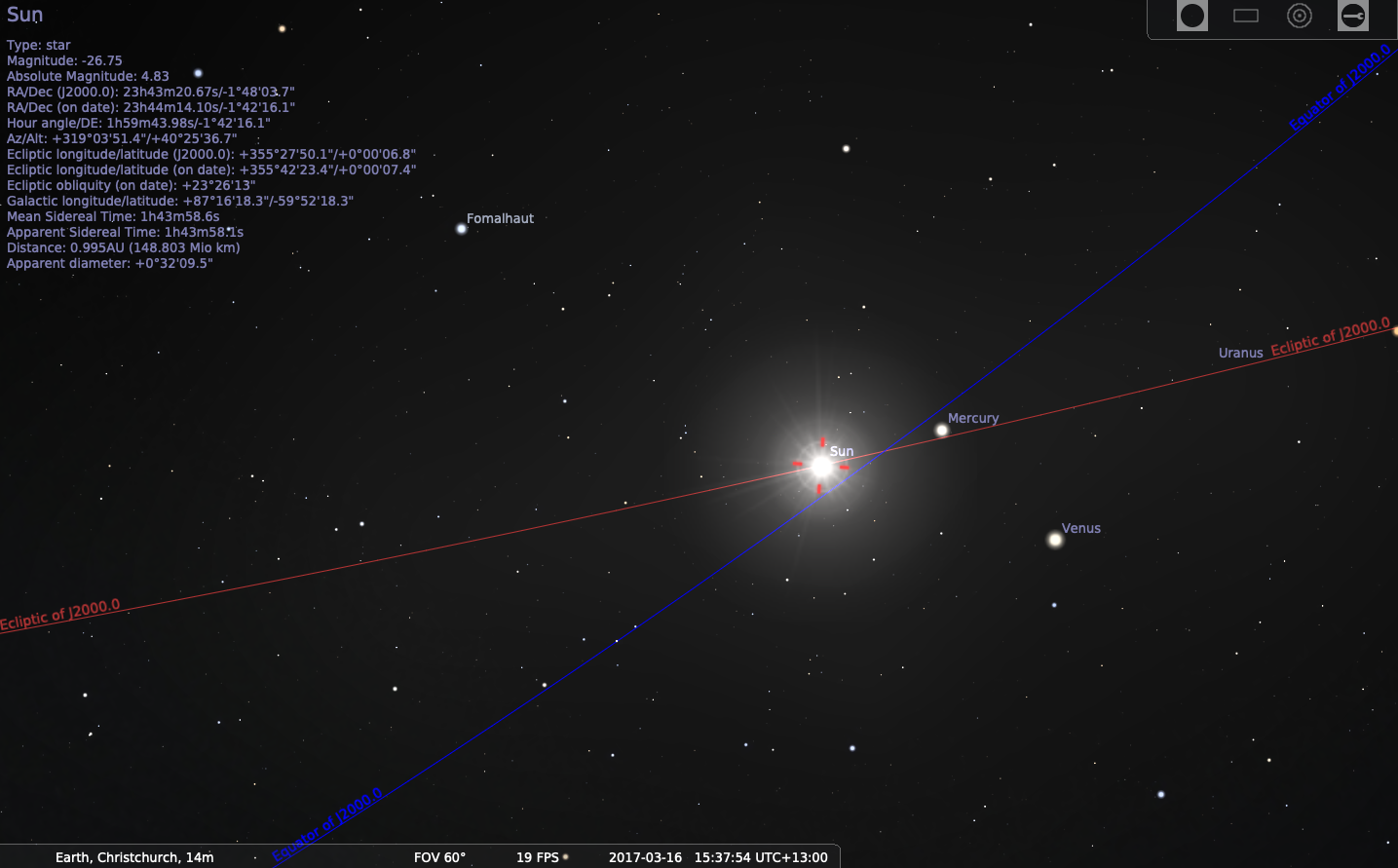}
\caption{(a) shows different trajectories corresponding to the obtained jerk-parallax solutions. The $red$ trajectory corresponds to P1, the $blue$ corresponds to P2, the $black$ corresponds to P3, and the $magenta$ corresponds to P4. (b) a screenshot from Stellarium software shows the Sun's path projected in the plane of the sky during the OGLE-2017-BLG-103's peak. The Sun's projected acceleration is along the ecliptic towards the intersection of both planes. The jerk velocity is $\sim${165.62 km/s}.}
\label{fig:4.3}
\end{figure*}

\noindent The single lens model did not fit the rising side of the asymmetric light curves of these events. So, we explore the orbital parallax model by adding two parameters $\pi_{EE}$ and $\pi_{EN}$. We initially perform a coarse grid search minimizing $\chi^{2}$ in the ($\pi_{EE}$, $\pi_{EN}$) plane by holding the Paczy\'nski parameters constant. For OGLE-2017-BLG-0103, we found a minima at ($\pi_{EE},\pi_{EN}$) = (0.184, -0.168) and for OGLE-2017-BLG-0192 at (0.120, -0.392) respectively. We seed these parameters along with Paczy\'nski parameters to $emcee$ \citep{foreman2013} and obtained a fitting model called P1. { We then re-normalized the photometric uncertainties in magnitudes using the P1 model so that ${\chi^{2}/d.o.f.}{\sim}$1.} After re-normalizing the photometric uncertainties, the P1 model was re-run, which converged to a solution shown in Figure (\ref{fig:triangle_both}a) where we see the  Jerk-parallax degeneracy in the discrete contours of the model parameters. Similarly, Figure (\ref{fig:triangle_both}b) shows the P1 model parameter distributions for OGLE-2017-BLG-0192 where we do not see any bimodal distributions of $\pi_{EN}$, $t_{E}$ and $t_{0}$ parameters which indicate that the degeneracy is broken in this case.\\
\noindent For OGLE-2017-BLG-0103, we term the model corresponding to the left mode in the $\pi_{EN}$ distribution as P3. To explore the mirrored degenerate solutions, which are also the ecliptic degenerate solutions, we seed the emcee with the P1 model but with the sign of $u_{0}$ reversed (see \cite{jiang2004, Poindexter2005}, \cite{skowron2011}). The walkers then converge to new solutions P2 and P4. Thus, we find all four degenerate solutions where P3 and P4 are the Jerk-Parallax degenerate solutions corresponding to P1 and P2, respectively. { While P4 is the constant acceleration degenerate pair of P3, it is the ecliptic degenerate counterpart of P1 (and similarly, P2 is the constant-acceleration pair of P1 and ecliptic degenerate P3). The trajectories for these pairs take the path shown in Figure (\ref{fig:4.3}a)}. So, we must expect similar lens properties for each pair of ecliptic degenerate solutions but different for the Jerk-Parallax degenerate solution. In Figure (\ref{fig:4.3}b), we show the contours of $\pi_{E}$ of all the solutions. Although we do not see any bimodal distributions in the case of OGLE-2017-BLG-0192, we still repeat the above procedure and find that the discrete solutions are not very far from each other in the $\pi_{E}$ plane (see Figure (\ref{fig:4.4}b)).
\subsection{Alternative method to find jerk parallax solutions}\label{section:alternate}
\begin{figure*}[!thtb]
  \centering
\includegraphics[width=0.45\textwidth]{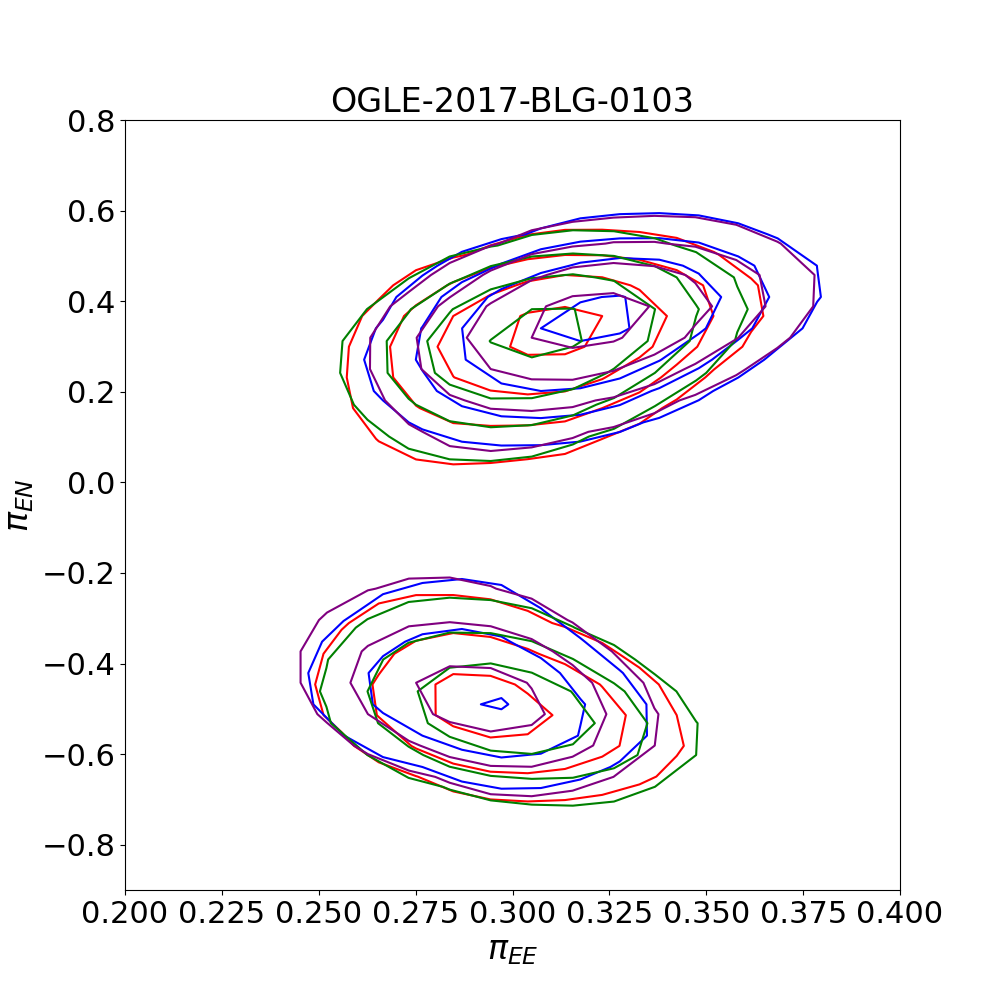}
\includegraphics[width=0.45\textwidth]{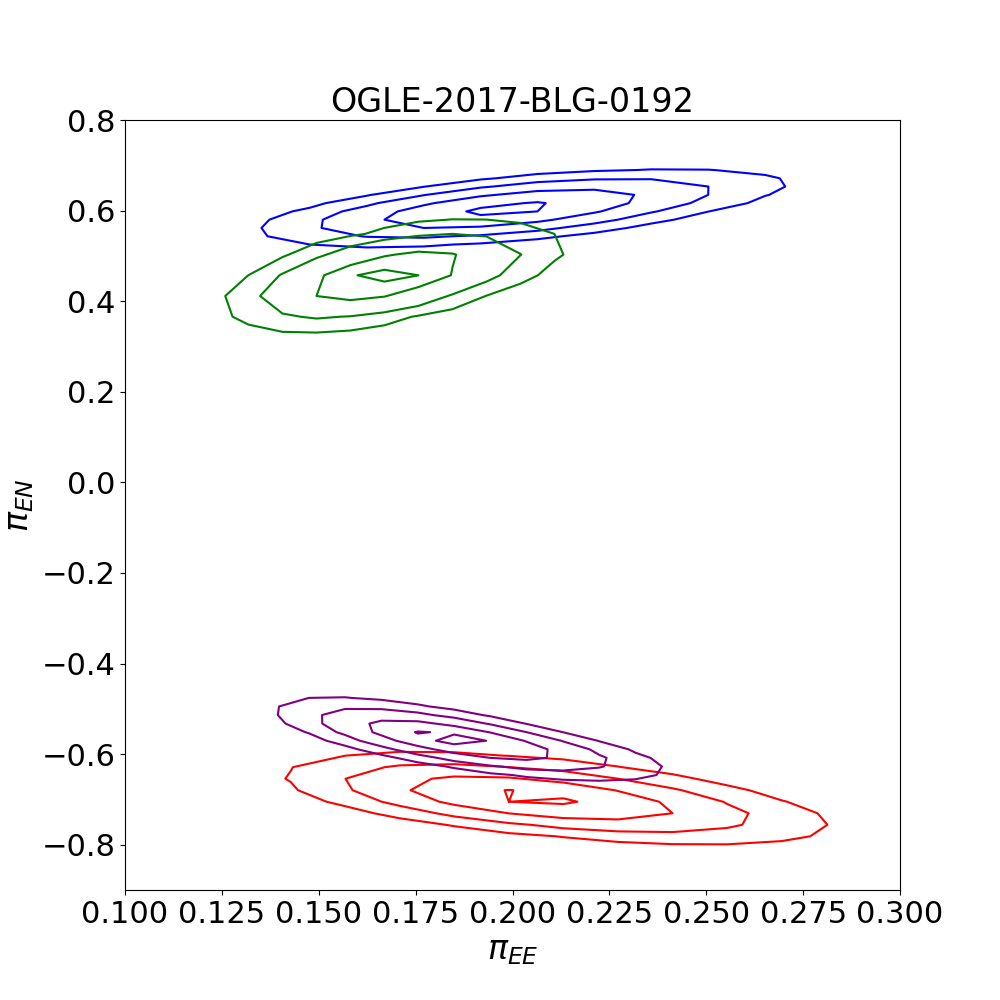}
\caption{In the first plot, the fit to the light curve with the orbital parallax model is red, while the black dotted model represents a Paczy\`nski Curve. The other two plots are the triangle plots showing the degenerate solutions. The right mode of the bimodal distribution of $\pi_{EN}$, $t_{E}$, and $t_{0}$ in the center plot shows the P1 model while the corresponding Jerk-Parallax model P3 is the left mode. Similarly, the solution P4 is the one that corresponds to the left mode of the bimodal distribution of $\pi_{EN}$, $t_{E}$, and $t_{0}$, while P2 is on the right. (P1,P2) and (P3,P4) are related by the $ecliptic$ degeneracy.}
\label{fig:4.4}
\end{figure*}

\begin{figure}
\centering
\includegraphics[width=0.47\textwidth]{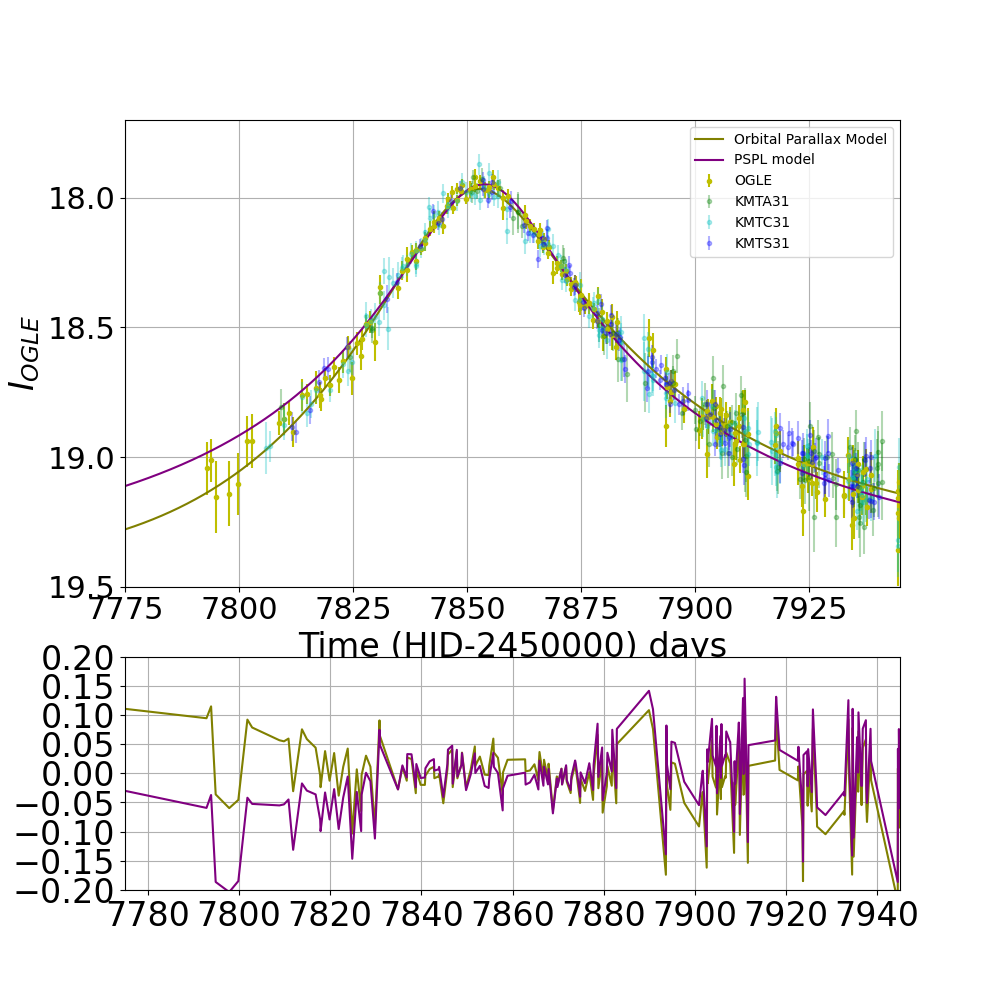}
\caption{Light curve of the event OGLE-2017-BLG-0192}
\label{fig:0192}
\end{figure}

\noindent We show an alternative way to find the Jerk-Parallax degenerate solutions for the event OGLE-2017-BLG-0103 based on the framework by \cite{an2002, smith2003a, gould2004} and also used by \cite{jiang2004} and \cite{park2004}. In this method, we find the Sun's celestial position at the event's peak and obtain the velocity ($v_{\oplus}$) and acceleration ($\alpha$). At the time of the vernal equinox of the year 2017 (which was close to $t_{0}$ of OGLE-2017-BLG-0103), the acceleration of the Sun projected into the sky was
\begin{equation*}
\alpha(n,e) = (-2.44\times 10^{-7}, 1.49\times 10^{-4})Km/day^{2}
\label{eq:4.3}
\end{equation*}
\noindent The projected acceleration points $359.80^{0}$ north through east in the sky. Since the parallax contours are elongated in the direction perpendicular to this direction at $269.9^{0}$, to find the model P3, we first find $\pi_{E,\parallel}$ and $\pi_{E,\perp}$, the components of $\pi_{E}$ parallel and perpendicular to the acceleration of the Sun in the sky at $\sim$ (0.43, 0.00). Then, the perpendicular component of the degenerate parallax solution is found by using the relation:
\begin{equation}
\pi'_{E,\perp} = -(\pi_{E,\perp}+\pi_{j,\perp})
\label{eq:1.30}
\end{equation}
\noindent where $\pi_{j,\perp}$ is the perpendicular component of the Jerk-Parallax and is calculated by,
\begin{equation}
\pi_{j,\perp} = -\frac{4}{3}\frac{1yr}{2\pi{t_{E}}}\frac{sin{\beta_{ecl}}}{(cos^{2}{\psi}sin^{2}{\beta_{ecl}} + sin^{2}{\psi})^{3/2}}
\label{eq:1.31}
\end{equation}
where $\beta_{ecl}$ is the ecliptic latitude of the event and $\psi$ is the phase of the Earth's orbit. \noindent Using Equation (\ref{eq:1.31}), we get $\pi_{j,\perp}$ = 0.17. Thus, the location of the degenerate parallax solution is: ($\pi_{E,\parallel}^{'},\pi_{E,\perp}^{'}$) = (0.43, -0.17). The Sun's path in the plane of the sky can be visualized in Figure (\ref{fig:4.3}b) where the background image is a screenshot of \textit{Stellarium} v.0.15.1\footnote{www.stellarium.org} and the ecliptic and equatorial planes are shown in red and blue colors, respectively. { We can also see the projected position of the Sun close to the vernal equinox, the direction of projected acceleration towards the intersection of the ecliptic and equatorial planes. There is one component of $\pi_{E}$ parallel to the acceleration along the ecliptic and two anti-parallel perpendicular components causing this degeneracy}.\\ 
\noindent In Figure (\ref{fig:4.4}a), we see that the major axes of the elliptical co-variances in the first plot for OGLE are nearly perpendicular to each other. Since the direction of {$\mathbf{\pi_{E}}$} is the direction of {$\mathbf{\mu_{rel}}$} in the adopted frame of reference, therefore, the direction of {$\pmb{\mu}_{rel}$} for P3 will be perpendicular to the direction of {$\pmb{\mu}_{rel}$} in P1 solution (see Figure (\ref{fig:4.3}a). As the direction of {$\pmb{\mu}_{rel}$} for P1 solution is arctan($\pi_{EN}$/$\pi_{EE}$) = $47^{0}$, the direction of {$\pmb{\mu}_{rel}$} for P3 solution should be $137^{0}$. Thus, we can convert ($\pi_{E,\parallel}$,$\pi_{E,\perp}'$) derived earlier to celestial frame ($\pi_{EE},\pi_{EN}$) = (0.30, -0.34) using \citep{an2002}. After seeding emcee with the Paczy\'nski parameters of the P1 solution and these parallax parameters, we found that the median value of the degenerate parallax parameters is ($\pi_{EE}, \pi_{EN}$) = (0.30,-0.50). Furthermore, $t_{0}$ and $t_{E}$ also converge to the other mode of the bimodal distribution of the samples. Similarly, the other P4 solution was found using the P2 solution. In all the solutions, it is observed that $\pi_{E,\parallel} > \pi_{E,\perp}$\footnote{{ $\pi_{E,\parallel}$ is responsible for the asymmetry about the base in the lightcurve. In contrast, $\pi_{E,\perp}$ is responsible for the asymmetry about the peak of the lightcurve.}} which is also seen in the light curve. The samples of the model parameters corresponding to each solution with their $16^{th}$, $50^{th}$, and $84^{th}$ percentile values are shown in Table (\ref{table:4.2}). Similarly, the model parameters for the different parallax solutions with their $16^{th}$, $50^{th}$ and $84^{th}$ percentile values for OGLE-2017-BLG-0192 are shown in Table (\ref{table:4.2o}) and the light curve in Figure (\ref{fig:0192}).
\begin{table*}[htbp]
\centering
\tiny
\setlength{\tabcolsep}{1pt}
\begin{tabular}{|p{1.4cm}|p{1.8cm}|p{1.8cm}|p{1.8cm}|p{1.8cm}|p{1.8cm}|p{1.8cm}|p{2.2cm}|}
\hline
\multicolumn{8}{|c|}{Microlensing model results} \\
\hline
Parameter & PSPL & P1 & P2 & P3 & P4 & Xallarap & Xallarap + Parallax \\
\hline
$\chi^2$ & 49123.42 & 11288.62 & 11290.50 & 11289.86 & 11290.90 & 11282.85 & 11237.26 \\
$u_{0}$ & 0.099 $\pm$ 0.001 & 0.123 $\pm$ 0.001 & -0.122 $\pm$ 0.001 & 0.123 $\pm$ 0.001 & -0.123 $\pm$ 0.001 & 0.104 $\pm$ 0.09 & $-0.144^{+0.006}_{0.008}$ \\
$t_{0}$ (days) & 7829.90 $\pm$ 0.01 & 7829.44 $\pm$ 0.09 & 7829.67 $\pm$ 0.03 & 7830.05 $\pm$ 0.06 & 7829.67 $\pm$ 0.03 & 7829.66 $\pm$ 0.06 & 7830.37 $\pm$ 0.04 \\
$t_{E}$ (days) & 73.12 $\pm$ 0.01 & 58.26 $\pm$ 1.62 & 60.42 $\pm$ 2.12 & 63.91 $\pm$ 2.22 & 60.23 $\pm$ 2.06 & 73.17 $\pm$ 3.22 & 53.63$\pm$2.07 \\
$\pi_{E,E}$ & - & 0.31 $\pm$ 0.02 & 0.30 $\pm$ 0.02 & 0.30 $\pm$ 0.02 & 0.30 $\pm$ 0.02 & - & -0.014 $\pm$ 0.004 \\
$\pi_{E,N}$ & - & 0.32 $\pm$ 0.08 & -0.02 $\pm$ 0.41 & -0.50 $\pm$ 0.11 & 0.02 $\pm$ 0.40 & - & 0.002 $\pm$ 0.002 \\
$\xi_{1}$ & - & - & - & - & - & -0.001 $\pm$ 0.002 & 0.001 $\pm$ 0.001 \\
$\xi_{2}$ & - & - & - & - & - & 4.28 $\pm$ 1.22 & $0.16^{+0.02}_{-0.01}$ \\
$\omega$ & - & - & - & - & - & 0.004 $\pm$ 0.001 & 0.024 $\pm$ 0.02 \\
$i$ & - & - & - & - & - & 1.56 $\pm$ 0.02 & 1.56 $\pm$ 0.01 \\
$\phi_{0}$ & - & - & - & - & - & 1.42 $\pm$ 0.14 & 1.74 $\pm$ 0.19 \\
$q_{s}$ & - & - & - & - & - & 0.0 & $0.51^{+0.03}_{-0.04}$ \\
$f_{s,\text{OGLE}}$ & 2135.62 $\pm$ 7.45 & 2142.26 $\pm$ 8.57 & 2142.10 $\pm$ 8.57 & 2139.82 $\pm$ 3.21 & 2159.93 $\pm$ 8.64 & 2140.78 $\pm$ 9.32 & 2140.78 $\pm$ 9.32 \\
$f_{s,\text{KMTA31}}$ & 1565.32 $\pm$ 9.80 & 1570.63 $\pm$ 11.40 & 1571.00 $\pm$ 11.41 & 1568.81 $\pm$ 5.49 & 1584.84 $\pm$ 11.51 & 1570.67 $\pm$ 11.12 & 1570.67 $\pm$ 11.12 \\
$f_{s,\text{KMTC31}}$ & 2275.56 $\pm$ 4.26 & 2323.37 $\pm$ 4.53 & 2323.60 $\pm$ 4.53 & 2321.11 $\pm$ 7.97 & 2343.57 $\pm$ 9.12 & 2321.25 $\pm$ 4.62 & 2321.25 $\pm$ 4.62 \\
$f_{s,\text{KMTS31}}$ & 2127.67 $\pm$ 7.38 & 2125.29 $\pm$ 7.37 & 2125.41 $\pm$ 7.37 & 2123.11 $\pm$ 7.37 & 2143.57 $\pm$ 7.44 & 2124.68 $\pm$ 7.03 & 2124.68 $\pm$ 7.03 \\
$f_{s,\text{KMTA42}}$ & 1570.51 $\pm$ 6.97 & 1578.11 $\pm$ 7.98 & 1578.17 $\pm$ 7.98 & 1575.87 $\pm$ 4.52 & 1591.36 $\pm$ 8.04 & 1576.63 $\pm$ 7.99 & 1576.63 $\pm$ 7.99 \\
$f_{s,\text{KMTC42}}$ & 2342.34 $\pm$ 4.57 & 2323.30 $\pm$ 5.49 & 2323.09 $\pm$ 5.49 & 2320.80 $\pm$ 11.39 & 2342.46 $\pm$ 5.54 & 2322.00 $\pm$ 5.30 & 2322.00 $\pm$ 5.30 \\
$f_{s,\text{KMTS42}}$ & 2126.89 $\pm$ 5.67 & 2125.77 $\pm$ 5.84 & 2125.38 $\pm$ 5.84 & 2123.36 $\pm$ 8.56 & 2143.01 $\pm$ 5.89 & 2126.87 $\pm$ 6.32 & 2126.87 $\pm$ 6.32 \\
$f_{b,\text{OGLE}}$ & 1801.27 $\pm$ 21.88 & 1803.28 $\pm$ 22.81 & 1810.18 $\pm$ 22.83 & 1811.93 $\pm$ 22.80 & 1800.28 $\pm$ 22.84 & 1801.21 $\pm$ 22.43 & 1563.44 $\pm$ 72.43 \\
$f_{b,\text{KMTA31}}$ & 2749.22 $\pm$ 11.20 & 2747.28 $\pm$ 11.43 & 2750.22 $\pm$ 18.63 & 2752.91 $\pm$ 11.43 & 2740.65 $\pm$ 11.51 & 2745.24 $\pm$ 12.21 & 2495.24 $\pm$ 180.21 \\
$f_{b,\text{KMTC31}}$ & 2212.63 $\pm$ 18.97 & 2213.61 $\pm$ 19.18 & 2219.03 $\pm$ 9.12 & 2221.69 $\pm$ 19.18 & 2205.86 $\pm$ 4.57 & 2745.24 $\pm$ 12.21 & 1833.24 $\pm$ 128.21 \\
$f_{b,\text{KMTS31}}$ & 12200.56 $\pm$ 16.03 & 12177.35 $\pm$ 15.99 & 12182.79 $\pm$ 16.01 & 12185.16 $\pm$ 15.99 & 12171.50 $\pm$ 19.22 & 12177.73 $\pm$ 16.78 & 11848.73 $\pm$ 316.78 \\
$f_{b,\text{KMTA42}}$ & 2735.54 $\pm$ 9.30 & 2738.75 $\pm$ 9.10 & 2743.07 $\pm$ 19.21 & 2745.75 $\pm$ 9.10 & 2735.71 $\pm$ 8.04 & 2736.98 $\pm$ 9.12 & 2519.98 $\pm$ 179.12 \\
$f_{b,\text{KMTC42}}$ & 2176.66 $\pm$ 15.98 & 2177.66 $\pm$ 18.60 & 2184.12 $\pm$ 13.92 & 2186.66 $\pm$ 18.60 & 2174.00 $\pm$ 5.54 & 2176.76 $\pm$ 17.76 & 1816.76 $\pm$ 127.76 \\
$f_{b,\text{KMTS42}}$ & 12175.72 $\pm$ 24.30 & 12172.96 $\pm$ 22.80 & 12180.16 $\pm$ 22.83 & 12181.64 $\pm$ 22.80 & 12170.82 $\pm$ 22.84 & 12175.96 $\pm$ 25.40 & 11859.96 $\pm$ 325.40 \\
\hline
\end{tabular}
\caption{The table shows the parameter means and uncertainties corresponding to each degenerate $\pi_{E}$ solution for the event OGLE-2017-BLG-0103. The corresponding linear fit parameters source flux ($f_{s}$) and the blend flux ($f_{b}$) with their uncertainties are also shown.}
\label{table:4.2}
\end{table*}

\begin{table*}[htbp]
\centering
\tiny
\setlength{\tabcolsep}{1pt}
\begin{tabular}{ |p{1.5cm} |p{2.0cm} |p{2.0cm} |p{2.0cm} |p{2.0cm} |p{2.0cm}|p{2.0cm}|}
\hline
\multicolumn{7}{|c|}{Microlensing model results} \\
\hline
Parameter & PSPL & P1 & P2 & P3 & P4 & Xallarap \\
\hline
$\chi^{2}$ & 4810.66 & 1540.73 & 1548.62 & 1581.93 & 1568.16 & 1636.83 \\
$u_{0}$ & 0.134 $\pm$ 0.005 & 0.14 $\pm$ 0.02 & -0.10 $\pm$ 0.02 & 0.08 $\pm$ 0.02 & -0.11 $\pm$ 0.02 & $-0.14^{+0.06}_{-0.07}$ \\
$t_{0}$ (days) & 7853.78 $\pm$ 0.06 & 7853.69 $\pm$ 0.19 & 7853.33 $\pm$ 0.19 & 7850.52 $\pm$ 0.39 & 7851.31 $\pm$ 0.29 & 7830.12$\pm$0.25\\
$t_{E}$ (days) & 122.70 $\pm$ 0.06 & 139.97 $\pm$ 22.16 & 196.42 $\pm$ 41.39 & 175.93 $\pm$ 30.66 & 137.01 $\pm$ 21.00 & $72.90^{+4.06}_{-5.90}$\\
$\pi_{E,E}$ & - & 0.21 $\pm$ 0.03 & 0.20 $\pm$ 0.03 & 0.17 $\pm$ 0.02 & 0.19 $\pm$ 0.02 & -\\
$\pi_{E,N}$ & - & -0.70 $\pm$ 0.05 & 0.61 $\pm$ 0.04 & 0.46 $\pm$ 0.04 & -0.57 $\pm$ 0.04 & -\\
$\xi_{1}$ & - & - & - & - & - & 0.0002 $\pm$ 0.0001 \\
$\xi_{2}$ & - & - & - & - & - & $2.36^{+0.58}_{-3.40}$ \\
$\omega$ & - & - & - & - & - & 0.0048 $\pm$ 0.0008 \\
i & - & - & - & - & - & 1.571 $\pm$ 0.001 \\
$\phi_{0}$ & - & - & - & - & - & $2.01^{+0.24}_{-1.47}$ \\
$\log(q_{s})$ & - & - & - & - & - & $-2.88^{+5.57}_{-4.79}$ \\
$f_{s,OGLE}$ & 1373.28 $\pm$ 5.88 & 1374.07 $\pm$ 6.00 & 1371.29 $\pm$ 6.01 & 1372.89 $\pm$ 6.00 & 1373.28 $\pm$ 6.02 & 2949.99 $\pm$ 14.70 \\
$f_{s,KMTA02}$ & 19738.09 $\pm$ 311.20 & 19737.93 $\pm$ 316.31 & 19732.00 $\pm$ 318.63 & 19735.91 $\pm$ 312.43 & 19735.65 $\pm$ 311.51 & 35827.25 $\pm$ 150.20 \\
$f_{s,KMTC02}$ & 11618.32 $\pm$ 102.97 & 11615.34 $\pm$ 102.09 & 11609.99 $\pm$ 103.12 & 11612.32 $\pm$ 103.18 & 11612.86 $\pm$ 102.57 & 21038.56 $\pm$ 198.67 \\
$f_{s,KMTS02}$ & 23098.56 $\pm$ 191.52 & 23099.35 $\pm$ 190.98 & 23104.18 $\pm$ 196.01 & 23085.16 $\pm$ 195.99 & 23071.50 $\pm$ 193.22 & 42085.30 $\pm$  22.08\\
$f_{b,OGLE}$ & 1308.62 $\pm$ 12.45 & 1308.13 $\pm$ 12.37 & 1306.06 $\pm$ 8.57 & 1306.53 $\pm$ 6.21 & 1309.65 $\pm$ 7.64 & -284.13 $\pm$ 7.89 \\
$f_{b,KMTA02}$ & 15652.32 $\pm$ 653.80 & 15648.00 $\pm$ 653.16 & 15711.00 $\pm$ 614.41 & 15648.81 $\pm$ 648.49 & 15804.84 $\pm$ 611.51 & 126.39 $\pm$ 4.56 \\
$f_{b,KMTC02}$ & 13201.72 $\pm$ 294.26 & 13156.13 $\pm$ 293.72 & 13178.62 $\pm$ 288.58 & 13165.11 $\pm$ 297.97 & 13149.57 $\pm$ 290.12 & 4229.28 $\pm$ 38.32 \\
$f_{b,KMTS02}$ & 18078.63 $\pm$ 460.59 & 18077.67 $\pm$ 467.27 & 18069.81 $\pm$ 480.01 & 18075.00 $\pm$ 477.37 & 18070.00 $\pm$ 467.44 & 159.63 $\pm$ 7.79\\
\hline
\end{tabular}
\caption{{ The table shows the parameter means and uncertainties corresponding to the fitted orbital parallax and xallarap model of OGLE-2017-BLG-0192. The corresponding linear fit parameters source flux ($f_{s}$) and the blend flux ($f_{b}$) with their uncertainties are also shown.}}
\label{table:4.2o}
\end{table*}

\subsection{Xallarap effect}\label{section:xallarap}
\begin{figure*}
\centering
\includegraphics[width=0.45\textwidth]{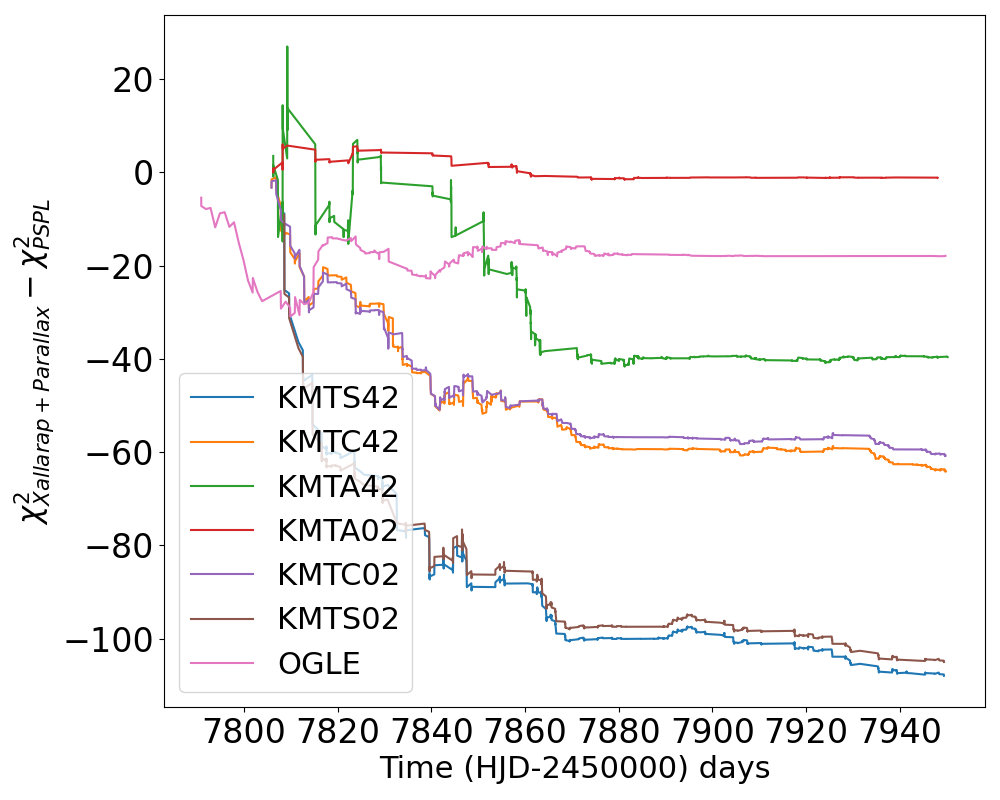}
\includegraphics[width=0.45\textwidth]{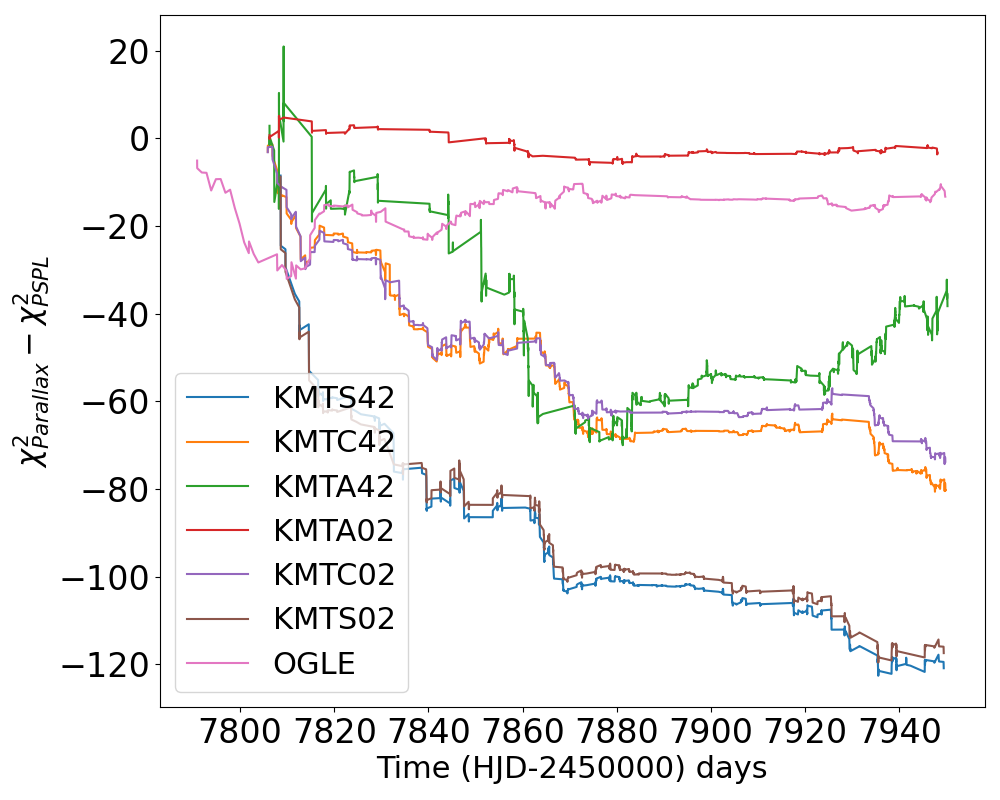}\\

\includegraphics[width=0.45\textwidth]{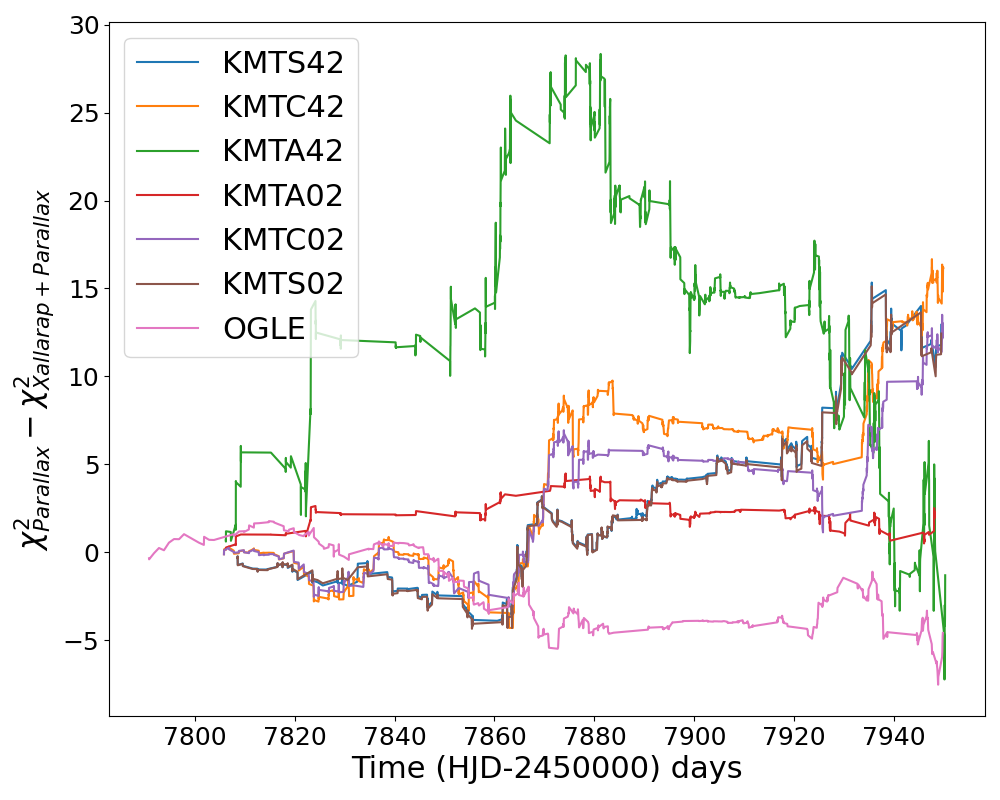}
\includegraphics[width=0.45\textwidth]{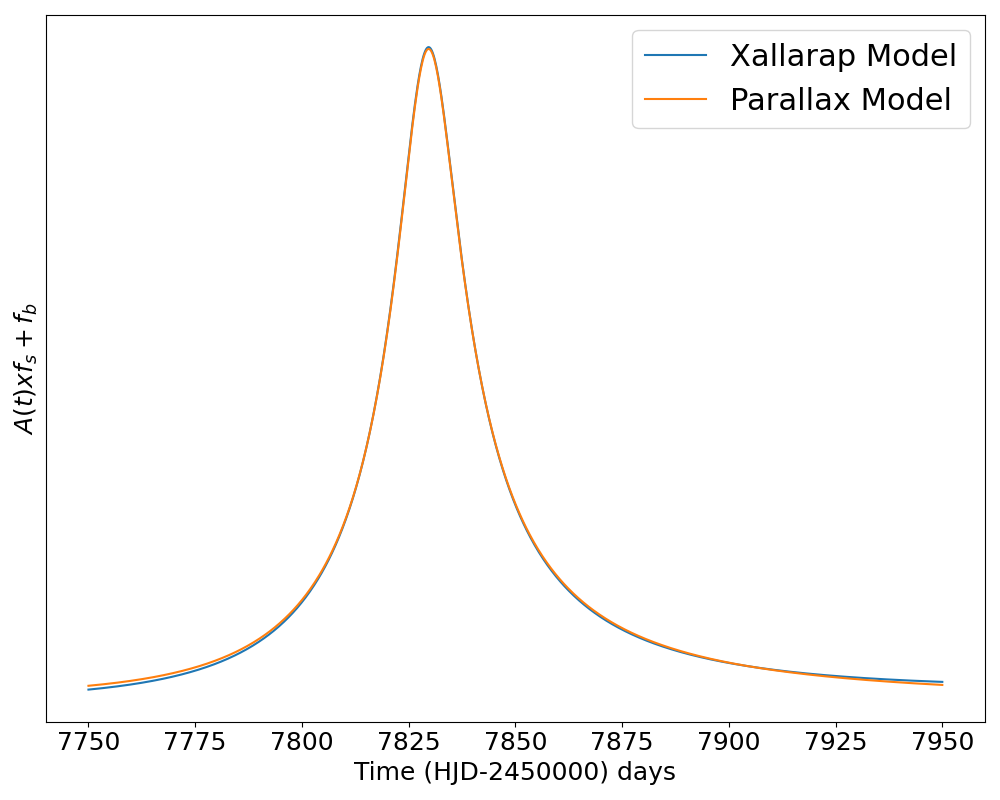}
\caption{{ In the figure above, the first three plots show the cumulative sum of the difference in the $\chi^{2}$ of (a) the Xallarap + Parallax model and the PSPL model, (b) the Parallax model and the PSPL model and (c) Xallarap Model and the PSPL model. The Parallax and Xallarap+Parallax models better fit than the PSPL model. We also see that the combined Xallarap + Parallax model shows a marked improvement over the Parallax model alone, as indicated by the positive $\Delta{\chi^{2}}$ values for most data sets KMT-A42 particularly. The OGLE dataset does not show significant deviation. So, while the Parallax model gives a better fit than the PSPL model, adding the Xallarap effect further improves the fit. In the last plot (d), we have shown the Xallarap+Parallax model (blue curve) plotted on top of the Parallax model (orange curve).}}
\label{fig:cumsum}
\end{figure*}

\begin{figure*} 
\centering
\includegraphics[width=0.47\textwidth]{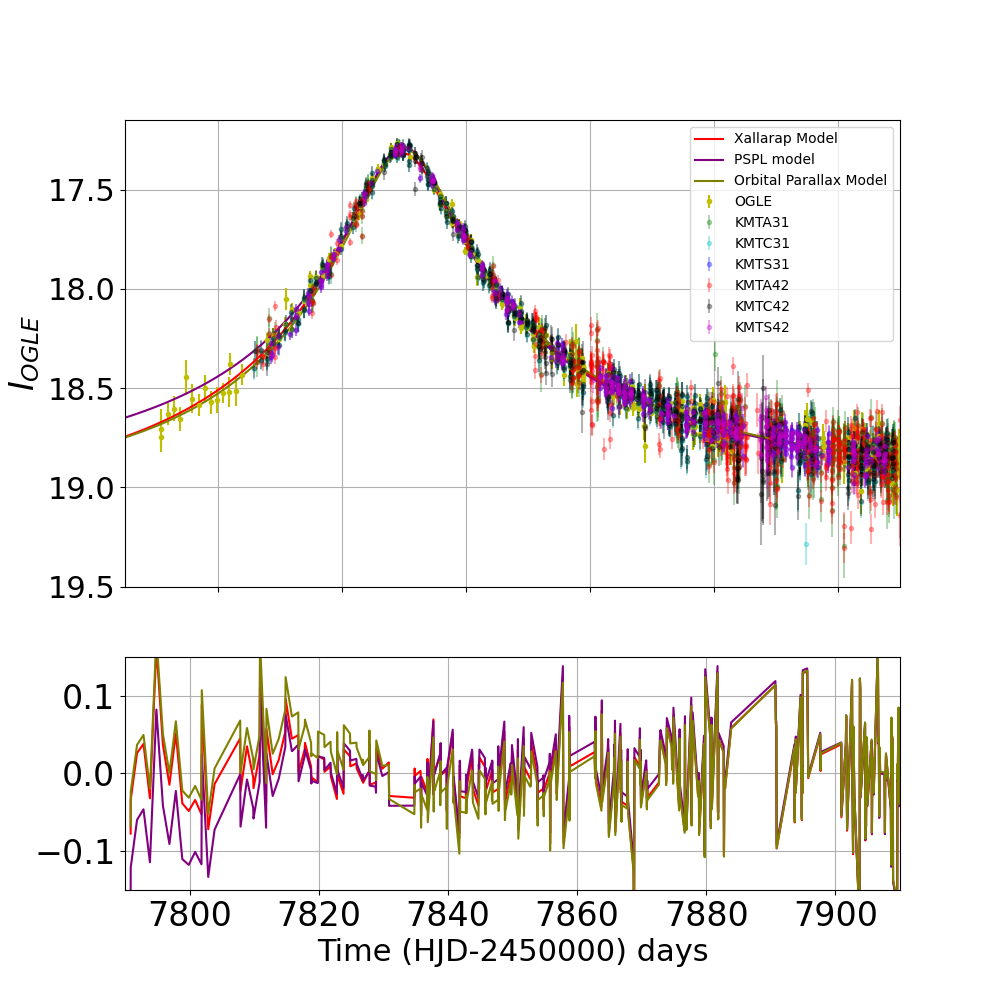}
\includegraphics[width=0.45\textwidth]{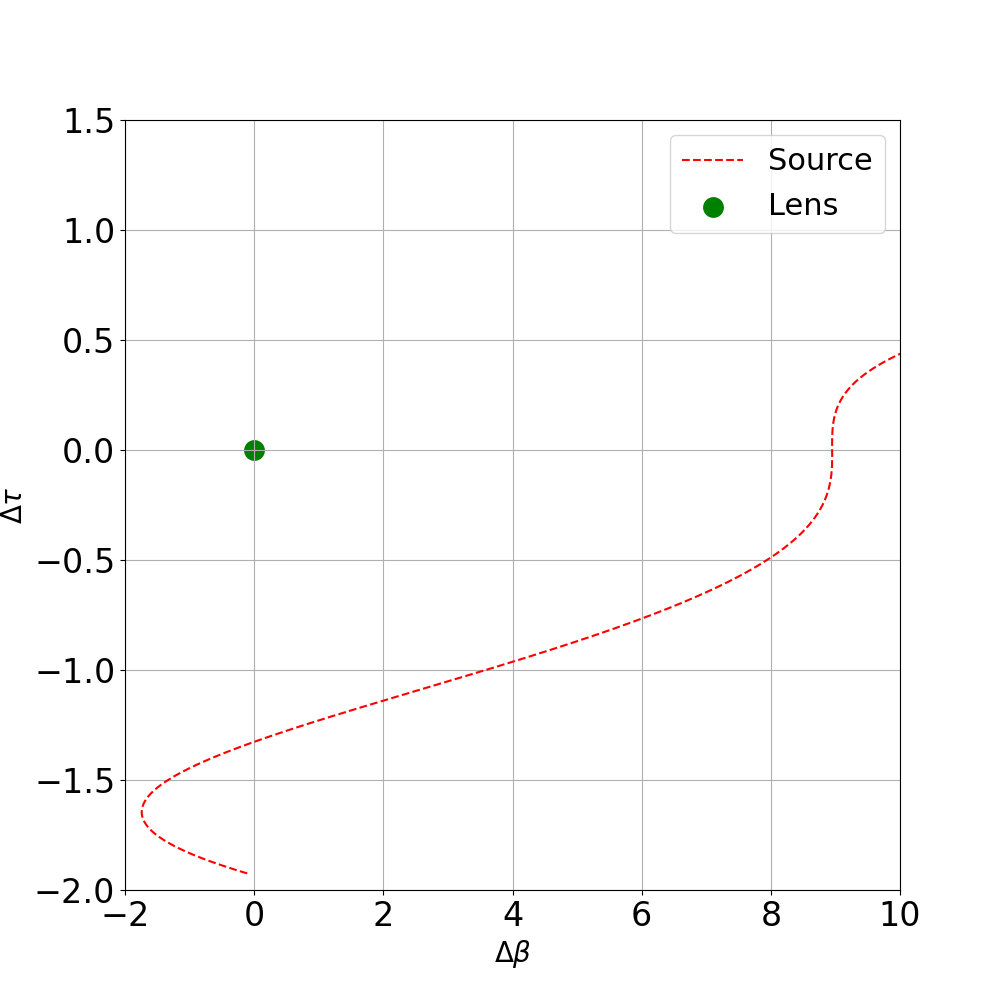}\\
\includegraphics[width=0.75\textwidth]{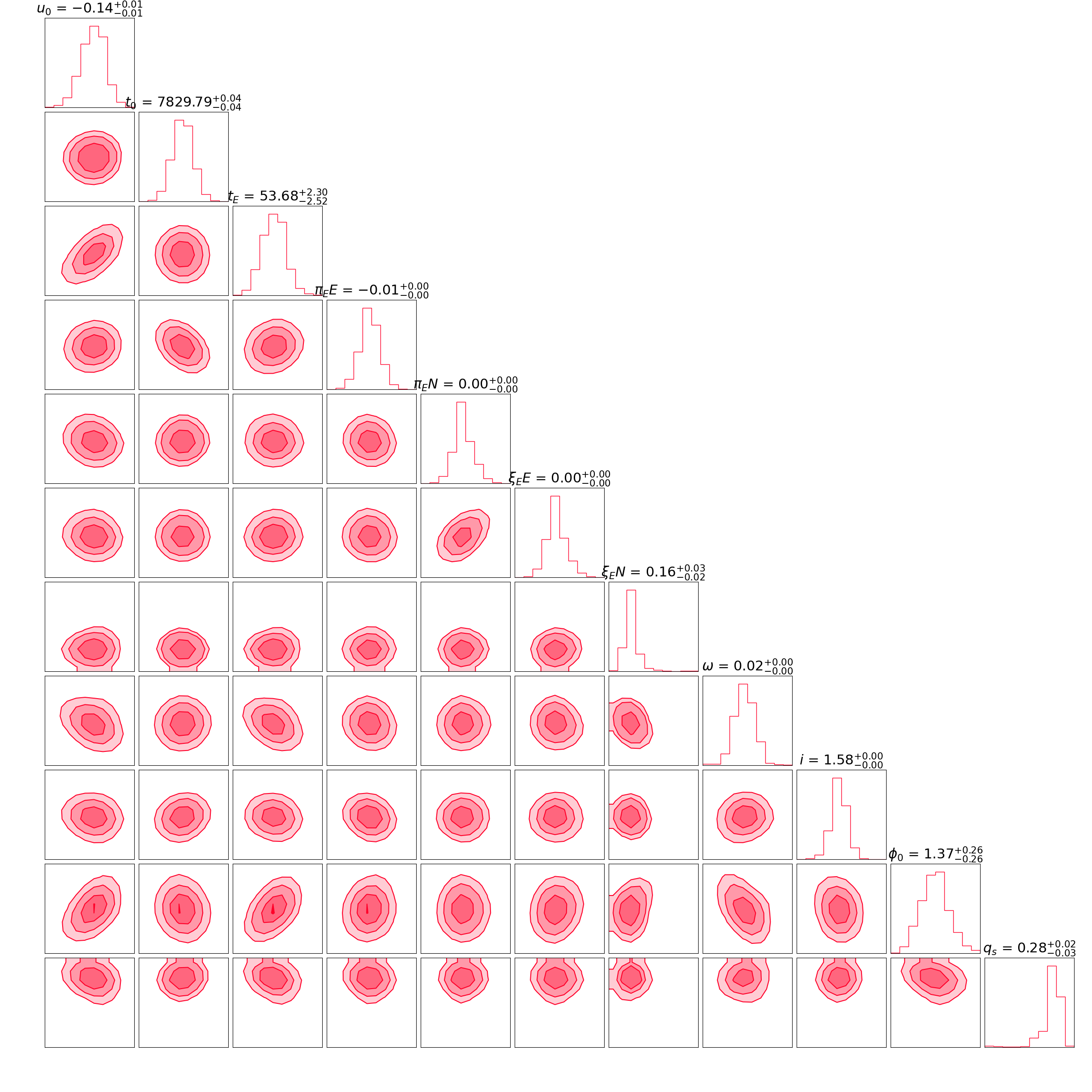}
\caption{{ The {\it first} plot shows the $PSPL$, $Orbital$ $Parallax$ and $Xallarap$+ $Parallax$ models fitted to the lightcurve of OGLE-2017-BLG-0103. The {\it second} plot shows the trajectory of the source system on the lens plane in the Xallarap + Parallax model, and the {\it last} plot shows the covariance of the Xallarap + Parallax model parameters. The values of these parameters at $16^{th}$, $50^{th}$ and $84^{th}$ quantiles are also given.}}
\label{fig:xallarap_trajectory}
\end{figure*}

\begin{figure}
    \includegraphics[width=0.45\textwidth]{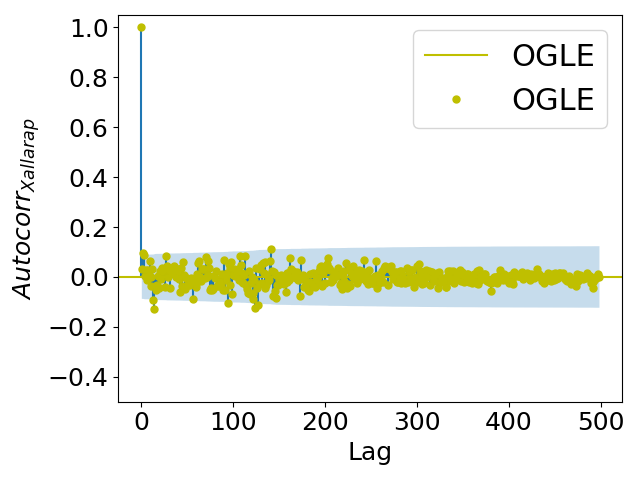}
    \includegraphics[width=0.45\textwidth]{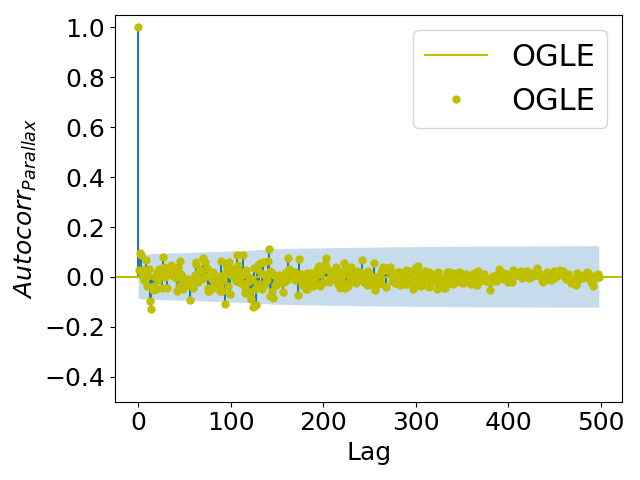}
    \caption{Autocorrelation of OGLE}
    \label{fig:autocorr_xalla}
\end{figure}

 \noindent Like the motion of the Earth around the Sun, a companion { orbiting} to the source might also cause similar perturbations in the light curve due to orbital parallax \citep{griest1992, paczynski1997, dominik1998, smith2003a}. { When the orbital effect is insignificant, the light curve of a binary-source event can be assumed as the superposition of the lightcurves of static individual sources. Conversely, in cases where the orbital effect is substantial, it necessitates the incorporation of the orbital dynamics. The study of binary sources and their orbital motion in microlensing lightcurves may even provide new opportunities: it may help break degeneracies in microlensing \citep{griest1992, han1997, paczynski1997}, or it may open a new channel to discover planets orbiting around the source star \citep{rahavar2009, bagheri2019, miyazaki2021}. The periods of binary systems may range from a few hours for very close systems to hundreds of years for widely spaced pairs. But binary source microlensing is not sensitive to long-period orbital signatures in the lightcurve \citep{rahavar2009}}. Also, given the steepness of the luminosity-mass relation, in most cases, the secondary star might be much fainter than the primary. Consequently, though we might not detect flux from the companion in such cases, the companion may still be able to modulate its trajectory on the lens plane. For this reason, the binarity of the source may sometimes arise just from the perturbation of the motion of the host without any signs of additional light, and the light curve can be similar to that of the orbital parallax modulated light curve. { We, therefore, explore this possibility in the case of both events by analyzing the lightcurve for the Xallarap effect {Parallax spelled backward}. The Xallarap vector ($\xi_{E}$) is the ratio of the semi-major axis of the source ($a_{s}$) normalized to $\theta_{E}$ projected onto the source plane;}
 \begin{equation}
     \xi_{E} = \frac{a_{s}}{\theta_{E}D_{s}}
     \label{eq:xie1}
 \end{equation}
 { where $D_{s}$ is the source distance. We form the Xallarap model by introducing seven more parameters: $\xi_{1}$ and $\xi_{2}$, which are the components of the Xallarap} parallel and perpendicular to the source velocity at $t_{0}$, $i$, which is the inclination of the orbital plane i, $\phi_{0}$ which is the phase from the ascending node, $\omega$ which is the angular velocity (day\textsuperscript{-1}), and $q_{s}$ which is the mass ratio of the secondary to the primary. We assume circular orbit and a power law mass-luminosity relation for the main-sequence stars (as can be seen later in the paper) $q_{f}$ = $q_{s}^{4}$, where $q_{f}$ is the flux ratio \citep{carroll2017} { (as can be seen from the result Table (\ref{table:4.2}))}. We use the $VBBinaryLensing$ code originally written in C++ but recently integrated into Python using $pybind11$ \citep{bozza2021}.\\
{ We initially model the light curves with a single lens and binary source xallarap model. For OGLE-2017-BLG-0103, the $\chi^{2}$ improves by $\sim$ 6; for OGLE-2017-BLG-0192, the $\chi^{2}$ worsens by $\sim$ 60, but the orbital parameters also do not mimic Earth's orbit. This indicates that the Xallarap model overfits the systematics in the light curve of OGLE-2017-BLG-0192, and the event is not affected by the xallarap effect. But even OGLE-2017-BLG-0103 has only a marginal improvement of $\chi^{2}$.} Therefore, we include orbital parallax in our Xallarap model (now known as the Xallarap+Parallax model) for OGLE-2017-BLG-0103. This further reduces the $\chi^{2}$ $\sim$ 39. So, our Xallarap+Parallax model converges to a solution with a better $\chi^{2}$ of $\sim$ 45 than the orbital parallax model. { In Figure (\ref{fig:cumsum}), we verify the authenticity of the Xallarap signal by plotting the cumulative $\Delta{\chi^{2}}$ for three models: the Xallarap+Parallax model compared to the parallax model, the Xallarap+Parallax model compared to the PSPL model, and the orbital parallax model compared to the PSPL model. The Parallax and Xallarap+Parallax models better fit than the PSPL model. We also see that the combined Xallarap+Parallax model shows a marked improvement over the Parallax model alone, as indicated by the positive $\Delta{\chi^{2}}$ values for most data sets, particularly KMT-A42. The OGLE dataset does not show significant deviation. So, while the Parallax model gives a better fit than the PSPL model, adding the Xallarap effect further improves the fit.} \\
{ From the model parameters of the event OGLE-2017-BLG-0103 listed in Table (\ref{table:4.2}), we see that the orbital inclination is $\sim$ $90^{+1.30}_{-0.55}$ degrees, the phase angle of the orbit is $\sim$ $100^{+12.78}_{-9.87}$ and the period of the orbit is $\sim$ $0.69^{+0.04}_{-0.06}$ years. The value of the xallarap vector is $\sim$ $0.16^{+0.02}_{-0.01}$, and the orbital parallax also converges to a small value, indicating that the xallarap effect is present in the lightcurve}. If we get a xallarap solution miming the Earth's orbit, the orbital parallax model is favored \citep{rahavar2009, miyazaki2021}. However, our results indicate that we are getting a different orbit that does not reflect Earth's orbit projected onto the source plane. { The mass ratio of the source system ($q_{s}$) is $0.28^{+0.03}_{-0.03}$, which indicates that the companion is not a planetary object.} As these values do not mimic the Earth's orbit, the xallarap solution cannot be ruled out. We show the lightcurve of OGLE-2017-BLG-0103 and the PSPL, Orbital Parallax and Xallarap + Parallax models fitted to it, the residuals plot, the trajectory of the source on the lens plane affected by the best-fit values of $\pi_{E}$ and $\xi_{E}$, and the covariance plots in Figure (\ref{fig:xallarap_trajectory}).\\
{ To check whether systematics are playing a role in Xallarap + Parallax model fit, we performed the autocorrelation analysis of the residuals eg. see Figure (\ref{fig:autocorr_xalla}) which shows the autocorrelation vs. time plot for the residuals in OGLE data\footnote{The plots for other datasets are shown in Appendix (\ref{appendix:autocorr}))}. In all the plots, the ACF at lag 0 is 1, as expected. The autocorrelation values are close to zero for all other lags and lie within the 95\% confidence interval (shaded blue region). This indicates that the residuals do not exhibit significant autocorrelation at any lag, suggesting the absence of systematic trends or correlated noise.  }
\section{Color-Magnitude Diagram for OGLE-2017-BLG-0103}\label{section:cmd}
\begin{figure}
\centering
\includegraphics[width=0.45\textwidth]{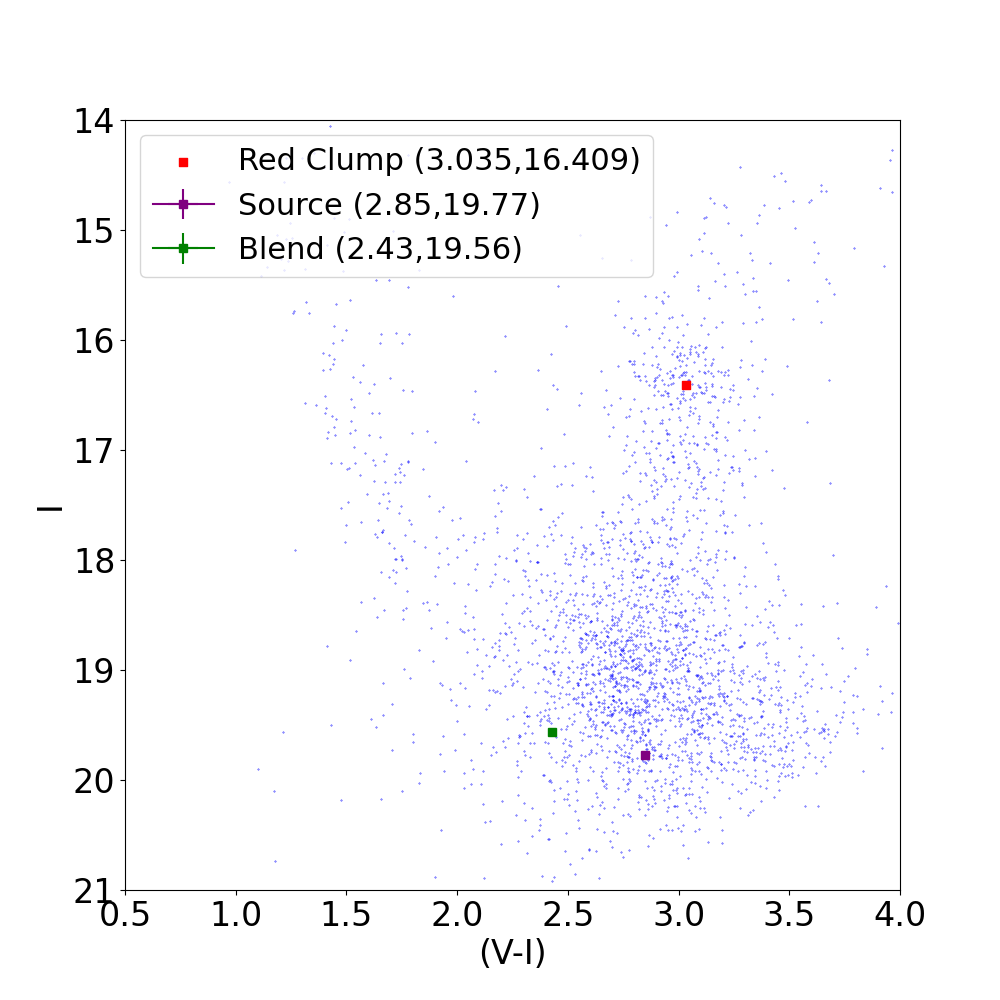}
\caption{The CMD of the stars in the field plotted using the V and I band data of KMTC BLG42. The centroid of the Red Clump stars, as estimated by pyDIA, is shown as a red-colored point, while the source and the blend from the lightcurve analysis of OGLE-2017-BLG-0103 are shown in the purple and green colors, respectively.}
\label{fig:cmd}
\end{figure}

\noindent We perform the source analysis only for OGLE-20187-BLG-0103 as we find xallarap signatures in the light curve. In the case of OGLE-2017-BLG-0192, we do not show the CMD here as there are no xallarap signatures in the light curve. Due to adequate coverage of this OGLE-2017-BLG-0103 by different observatories, the source and blend flux are fairly constrained. We then fit the V-band data of the KMTC-BLG42 field to our model and get the values of the source flux and the blend flux in the V-band as we got for the I-band. We use these fluxes to locate the position of the source and the blend on the CMD formed by the stars in the BLG02 field of the KMTC02 dataset (see Figure (\ref{fig:cmd})) and find that the source lies on the main-sequence. The position of the source and the blend on the CMD is { ($((V-I),I)$) = (2.85$\pm$0.04, 19.77$\pm$0.42), (2.43$\pm$0.03, 19.56$\pm$0.46) respectively}. In the CMD, the centroid of the red clump stars in the field, the source, and the blend are shown in red, purple, and green colors, respectively. We can see that the source lies on the main sequence, and the blend is located on the bluer side of the CMD but is slightly brighter than the source. We also find the source in the Gaia DR3 database \citep{gaia2022} with a source id of 4056392106235284992, source parallax ($\pi_{s}$), equatorial proper motion ($\mu_{ra}$, $\mu_{\delta{*}}$) = (-2.10 $\pm$ 0.73 mas, -8.34 $\pm$ 1.11 mas/yr, and -7.15$\pm$ 0.60 mas/yr) respectively. The re-normalized unit weight error ($ruwe$) of the source is 1.14, which indicates that the source is a single star or there is no bright companion. The negative value of parallax indicates that the source is faint and heavily blended, which is also evident from the lightcurve and the CMD.\\
From \cite{Nataf2013}, we find that the intrinsic color of the red clump in the bulge is at ${((V-I),I)}_{RC}$ = (1.06, 14.39). On comparing these values with the values for the centroid of the red clump from pyDIA, we estimate the reddening ($\Delta{V-I}$) and extinction $\Delta{I}$ towards the BLG02 field as (2.13,2.10) respectively. { This gives the true color and brightness of the source as $({(V-I), I)}_{0,s}$ = (0.71$\pm$0.04, 17.66$\pm$0.42)}.
\section{Lens properties}\label{seection:lens}
\noindent As mentioned earlier, the modeling of the microlensing light curve does not directly give information about the nature of the lens, e.g., its mass and distance, and it can be measured only if $\theta_{E}$ and $\pi_{E}$ are measured together from the lightcurve (see Equation (\ref{eq:1.20})). In the case of our events, we do not detect any finite source effects nor have any high-resolution imaging by the Hubble Space Telescope or interferometry observations by VLTI to measure $\theta_{E}$ directly. { So, we form a galactic model and simulate the microlensing events that could give rise to measured microlensing model parameters eg. $t_{E}$ and $\pi_{E}$ in the case of the Orbital Parallax model and $\xi_{E}$ in case of Xallarap+Orbital Parallax model}. We formulate the galactic model priors based on the Gaussian velocity distribution towards the bulge  \citep{han1995}, stellar density distribution, and the Chabrier stellar mass function (see \cite{chabrier2003, batista2011, clanton2014, zhu2017, jung2018} and others). We assume thin, thick discs with a normalizing factor of 0.12 \citep{juric2008, jung2018}. We assume the lens and source distances are free parameters in the galactic model. In addition, we assume certain different conditions to obtain the lens mass and distance using Orbital Parallax and Xallarap+Parallax models. Since we have obtained the best fit using the orbital parallax model for both events and another possibility using the Xallarap+Parallax model for OGLE-2017-BLG-0103, we discuss our results of the galactic model for both methods separately.\\
\subsection{Physical properties of the lens for both the events using only the Orbital Parallax model}
{ Here, we discuss the results using the orbital parallax model. Our galactic model is constrained by the samples of $\pi_{E}$ and $t_{E}$. Since the source star of OGLE-2017-BLG-0103 has proper motion information in Gaia Dr3, we also constrain our galactic model using this information. For this, we transform the proper motion in the galactic plane by referring to the relations by \cite{poleski2013}. However, we cannot constrain the galactic model for OGLE-2017-BLG-0192 as the source star has no astrometry information in Gaia Dr3. So, we use only the $t_{E}$ and $\pi_{E}$ information in this event. The results for OGLE-2017-BLG-0103 are given in Table (\ref{table:n10}) and the results corresponding to the least $\chi^{2}$ model of OGLE-2017-BLG-0192 in Table (\ref{table:n11}). These tables also show the lens's proper motion values and expected brightness (assuming it is a main sequence star).} The expected brightness includes the extinction towards the lens ($A_{I,L}$) computed by using,
\begin{equation}
    A_{i,L} = \frac{1-e[-D_{L}/h_{dust}]}{1-e[-D_{s}/h_{dust}]}A_{I,s}
\label{eq:ail}
\end{equation}
where $A_{I,s}$ is the extinction to the source (=$\Delta(I)$ found in the CMD analysis), and $h_{dust}$ is the scale height of the dust towards the line of sight (see \cite{bennett2015}). From the result table, we see that for OGLE-2017-BLG-0103, the P1 and P3 pair of solutions give similar results of a very low mass star having a typical $\mu_{re}$ of the disc stars while the P2 and P4 pair of solutions give similar results where the lens is more massive and has higher $\mu_{rel}$ than the disc stars. The expected brightness of these massive lenses computed assuming they lie on the main sequence is brighter than the blend, which is impossible. This indicates the lens might be a stellar-remnant and not a main-sequence star.
\subsection{Physical properties of lens and source for OGLE-2017-BLG-0103 using the Xallarap+Parallax model}
\begin{figure}
\centering
\includegraphics[width=0.45\textwidth]{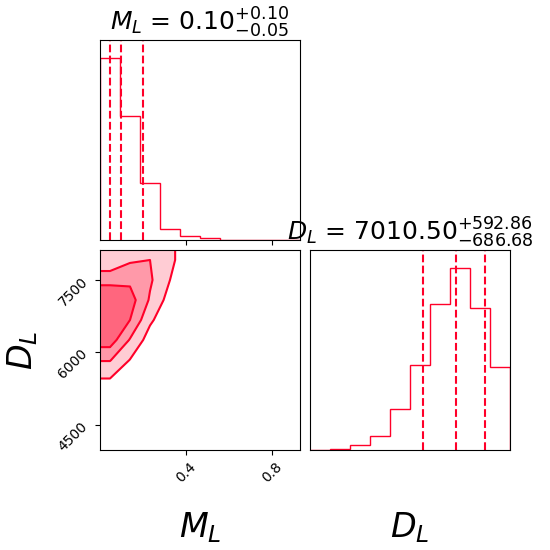}
\caption{{ The posterior distributions of $M_{L}$ ($M_{\odot}$), and $D_{L}$ ($kpc$) shown at $16^{th}$, $50^{th}$ and $84^{th}$ percentile levels.}}
\label{fig:gmx}
\end{figure}

\noindent { Our Xallarap+Parallax model consists of a single lens but a binary source. We estimate the source's mass ($M_{s}$) using {\it MESA Isochrones and Stellar Evolutionary Tracks (MIST) models} (see \citep{dotter2016,choi2016} for more information)\footnote{Also visit \url{http://waps.cfa.harvard.edu/MIST/index.html} to access the database and various tools.} its color and brightness information (calculated in the Section (\ref{section:cmd})) as 1.28$M_{\odot}$. Thus, using Kepler's third law, we get}
\begin{equation}
    a = \Big((M_{sh}+M_{sc})P^{2}\Big)^{1/3}
\label{eq:asc}
\end{equation}
{ where $M_{s}$ = $M_{sh}$ + $M_{sc}$, $P$ is the orbital period and $a$ is the semi-major axis of the orbit. As we know $\omega$ and $q_{s}$ from our Xallarap+Parallax model, we get $a$=0.98$\pm$0.12 A.U., $M_{sh}$ = $1.01^{+0.1}_{-0.2}$ $M_{\odot}$ and $M_{sc}$ = $0.27^{+0.04}_{-0.05}$ $M_{\odot}$. Now, $\xi_{E}$ is related to $\theta_{E}$ by}
\begin{equation}
\xi = \frac{1}{\theta_{E}{D_{s}}}P^{2/3}\frac{\Big(q_{s}M_{s,h}^{1/3}\Big)}{(1+q_{s})^{2/3}}
\label{eq:kep_thetae}
\end{equation}
{ Thus, we get $\theta_E$ = 0.16 $\pm$ 0.04 mas. Using the galactic model, we can constrain the mass of the lens and distance to it. In our galactic model analysis using the Xallarap+Parallax model, we keep $M_{L}$ and $D_{L}$ as free parameters such that $D_{L}$ is always less than $D_{s}$. We use $I_{b}$ as the upper limit on the lens brightness, assuming that the lens is also a main-sequence star. Since $\pi_{E}$ converges to a very small value in the Xallarap+Parallax model, we ignore the parallax constraint\footnote{We also do not use Equation (\ref{eq:1.20}) as it will lead to a massive lens.}. Instead, we use Equation (\ref{eq:kep_thetae}). We use the other priors similar to the ones used in the previous galactic model analysis using the Orbital Parallax model. We show the distribution of $M_{L}$ and $D_{L}$ in Figure (\ref{fig:gmx}). Interestingly, as the xallarap is more sensitive to the lenses closer to the source \citep{dominik1998}, the lens moves further and is closer to the source this time than in the earlier galactic model analysis.}
{ The physical parameters are listed in the last row of Table (\ref{table:n10})}.
\begin{table*}
\begin{center}
\setlength\tabcolsep{6.5pt}
\begin{tabular}{|p{1.6cm} |p{1.3cm} |p{1.3cm}|p{1.3cm}|p{1.3cm}|p{1.3cm}|p{1.2cm}|p{1.2cm}|p{1.1cm}|p{1.1cm}|}
\hline
{ Solution} & $\mathbf{D_{L}}$ (kpc) & $\mathbf{M_{L}}$ ($M_{\odot}$) & $\mathbf{D_{s}}$ (kpc) & $\mathbf{\mu_{rel}}$ (mas/yr) & $\mathbf{I_{L}}$ & $\mathbf{M_{s,h}}$ ($M_{\odot}$) & $\mathbf{M_{s,c}}$ ($M_{\odot}$) & $\mathbf{a}$ (A.U.) & $\mathbf{P}$ (years)\\
\hline
P1 & $1.98^{+0.79}_{-0.68}$ & $0.21^{+0.23}_{-0.16}$ & $7.81^{+0.39}_{-0.31}$ & $5.03^{+1.71}_{-1.25}$ & $26.48^{+4.38}_{-2.92}$ & - & - & - & -\\
\hline
P2 & $1.61^{+0.79}_{-0.63}$ & $0.67^{+0.45}_{-0.23}$ & $7.82^{+0.37}_{-0.30}$ & $9.16^{+3.22}_{-2.94}$ & $18.00^{+4.61}_{-2.68}$ & - & - & - & -\\
\hline
P3 & $1.92^{+0.79}_{-0.63}$ & $0.14^{+0.14}_{-0.06}$ & $7.81^{+0.38}_{-0.30}$ & $4.01^{+0.80}_{-0.71}$ & $27.16^{+6.87}_{-3.95}$ & - & - & - & -\\
\hline
P4 & $1.51^{+0.79}_{-0.63}$ & $0.70^{+0.24}_{-0.19}$ & $7.81^{+0.37}_{-0.30}$ & $10.58^{+3.49}_{-2.69}$ & $17.70^{+5.18}_{-3.64}$ & - & - & - & -\\
\hline
Xallarap + Parallax & $7.01^{+0.59}_{-0.68}$ & $0.10^{+0.10}_{-0.05}$ & 8.20 & $1.24^{+1.02}_{-0.98}$ & $27.82^{+2.93}_{-2.56}$ & $1.01^{+0.33}_{-0.25}$ & $0.27^{+0.12}_{-0.13}$ & $0.98^{+0.14}_{-0.12}$ & $0.86^{+0.04}_{-0.06}$\\
\hline
\end{tabular}
\caption{The result table shows the mass of the lens and the distance to it for all four Jerk-Parallax degenerate solutions - P1, P2, P3, and P4 of OGLE-2017-BLG-0103 obtained after performing the Bayesian analysis of the galactic model. P1 and P3, a pair of solutions, give similar values of $M_{L}$ and $D_{L}$, whereas P2 and P4 give similar values as expected from \cite{gould2004}. The brightness in the I-band is extinction corrected using Equation (\ref{eq:ail}) and is calculated using the power-law mass-luminosity relation, assuming that the lens is a main-sequence star. The last row shows the galactic model results when the Xallarap+Parallax model is considered.}
\label{table:n10}
\end{center}
\end{table*}

\begin{table*}
\centering
\footnotesize
\setlength\tabcolsep{3pt}
\begin{tabular}{|p{2.5cm} |p{2.5cm} |p{2.5cm}|p{2.5cm} |p{2.5cm}|p{2.5cm}|}
\hline
{Solution} & $\mathbf{D_{L}}$ {(pc)} & $\mathbf{M_{L}}$ {($M_{\odot}$)} & $\mathbf{D_{s}}$ {(pc)} & $\mathbf{\mu_{rel}}$ {(mas/yr)} & $\mathbf{I_{L}}$\\
\hline
P1 & $1796.05^{+799.89}_{-686.39}$ & $0.10^{+0.21}_{-0.07}$ & $7641.87^{+342.58}_{-304.96}$ & $1.54^{+0.50}_{-0.39}$ & $23.85^{+6.47}_{-5.11}$\\
\hline
\end{tabular}
\caption{The result table shows the mass of the lens, distance to the lens, relative lens-source proper motion, and the expected brightness in the I-band of the P1 solution obtained for OGLE-2019-BLG-0192, obtained after performing the Bayesian analysis of the galactic model. The brightness in the I-band is calculated using the power-law mass-luminosity relation, assuming that the lens is a main-sequence star.}
\label{table:n11}
\end{table*}
\section{Discussion}\label{section:discussion}
\noindent In this study, we have analyzed two microlensing events, OGLE-2017-BLG-0103 and OGLE-2017-BLG-0192, to try to identify the nature of the {\it microlens} and source. { These events peaked close to the vernal equinox in 2017 when the jerk-velocity ($v_{j}$) and the projected acceleration of the Sun ($\mathbf{\alpha}$) were very high. This introduces a Jerk-Parallax degeneracy in events with $t_{E}$ less than 60 days. While this degeneracy is manifested as continuous contours in the $\pi_{E}$ plane, for events with slightly longer $t_{E}$ and higher value of $\mathbf{\alpha}$, the degeneracy is converted to a discrete degeneracy (e.g., OGLE-2017-BLG-0103). For even larger $t_{E}$, we find that the high value of ${\mathbf{\alpha}}$ and $t_{E}$ splits the $\pi_{E}$ contours, and the degeneracy is resolved (e.g., OGLE-2017-BLG-0192).}\\
For OGLE-2017-BLG-0103, while we find that orbital parallax is responsible for the asymmetry in the lightcurve, we also find that the { xallarap effect added to the Orbital Parallax model also explains the asymmetry with an improved $\Delta{\chi^{2}}$ of $\sim$ 45. The source orbital parameters do not reflect the Earth's orbit, indicating that the model does not reflect Earth's orbit on the source plane. When plotted, the cumulative distribution of ${\chi^{2}_{Parallax}}$ - $\chi^{2}_{Xallarap+Parallax}$ plots shows an increasing trend indicating that the xallarap signal is present throughout the lightcurve and is improving the fit. As another quality check, we plotted the autocorrelation of the residuals of the Xallarap+Parallax model at various lags of all the data sets. We found that the correlation of the data is much less, thereby indicating no influence of the systematic trend in the data on the fitted model.} This indicates severe tension in both the Orbital Parallax and Xarallarap+Parallax models.\\
Previous works by \cite{dominik1998, smith2003a, smith2003b, rahavar2009, miyazaki2021} have shown that the xallarap signal is prominent for super-Jupiter mass planets in close orbits. This signal increases with the mass, and the wiggles in the light curve are most visible when the orbital period is small, and the planet's mass is higher (see \cite{rahavar2009, miyazaki2021}). 
In the case of OGLE-2017-BLG-0103, the companion is stellar, and the orbital period is ($\sim$ 0.8 years); hence, we see an asymmetry in the light curve instead of wiggles. Such events are expected to be common during the Nancy Grace Roman Space Telescope era \citep{spergel2015}, (which was previously named WFIRST) (see \cite{bennett2018a, penny2019, gaudi2019}.\\
{ To derive the source properties, we obtained each dataset's source and the blend fluxes by performing the linear regression of the microlensing model to the data.} We use the V-band data of KMTC02 to plot the CMD and then locate the position of the source and the blend on the CMD. The blend is bluer and slightly brighter than the source. Based on its position on the color-magnitude diagram (CMD), we find that the source is a main sequence star.\\
Given the inherent limitations of the microlensing light curve of both events, which hinder the direct measurement of finite source effects and, consequently, the Einstein radius ($\theta_{E}$), we turned to a galactic model analysis to estimate the mass and distance of the lens. { For OGLE-2017-BLG-0192, the galactic model analysis constrained by the samples of $t_{E}$ and $\pi_{E}$ of the best-fitted P1 solution gives $M_{L}$ $\sim$ 0.1 $M_{\odot}$ and located in the near disc. Assuming that the lens is a main-sequence star, we find that it is very faint ($I_{L}$ = 24 mag)}. For OGLE-2017-BLG-0103, we did two separate analyses for the Orbital Parallax and Xallarap+Parallax models. The galactic model for OGLE-2017-BLG-0192 was constrained only by the samples of $t_{E}$ and $\pi_{E}$. However, the galactic model for OGLE-2017-BLG-0103 was constrained not only by the samples of $t_{E}$ and $\xi_{E}$, but also the source proper motion ($\mu_{s}$) information from Gaia Dr3. The orbital parallax model gives us two similar solutions: one gives a very low-mass main-sequence star in the disc, and the other gives a more massive star and nearer than the former\footnote{as predicted by \citep{gould2004}, both pairs of Jerk-Parallax solutions give different properties for the lens.}. Assuming again that the lens is a main-sequence star, the expected I-band brightness of the lens for the former pair of solutions is $>$ 25 mag while it is $\sim$ 18 mag in case of later. If the later pair of solutions are correct, then this will imply that the lens is itself the blend and the lens-source pair can be observed using high-resolution telescopes by the year 2027 (considering the proper motion values given in Table (\ref{table:n10})).\\
The galactic model analysis using the { Xallarap+Parallax model} shows that the lens is $\sim$ 0.1$M_{\odot}$ dwarf star in the galactic bulge. As expected, our galactic model, constrained by the xallarap signal in the light curve, moves the lens further than the galactic model constrained by only orbital parallax. In our galactic model analysis, we have assumed the source to be located in the bulge at the distance of 8.2kpc. Using MESA isochrone models, we find that the source host has a mass $\sim$ 1.01$M_{\odot}$ and it is orbited by an $\sim$ 0.27$M_{\odot}$ dwarf-star at a separation of $\sim$ 0.98 A.U. and has an orbital period of $\sim$ 0.8 years.\\
\noindent Finally, the question arises: which model is true? Degeneracies are difficult to break in case of single-lens events due to the symmetry in the magnification pattern around the point caustic of the lens. If we apply Ocazm's razor to the event OGLE-2017-BLG-0103, the orbital parallax model is most likely as it is explained by fewer bodies and a smaller number of parameters. However, the degenerate pairs of orbital parallax solutions also have different lens properties. \cite{rahavar2009} state that an $\Delta{\chi^{2}}$ $\geq$ 11.09 is necessary to claim a xallarap detection. In our case, we have a $\Delta{\chi^{2}}$ of $\sim$ 45. Moreover, combined orbital parallax and xallarap signals show an increasing trend in the cumulative distribution of the $\Delta{\chi^{2}}$ than only the orbital parallax signal. We also do not find any influence from systematics. Indeed, for many microlensing events
of interest, models with xallarap may compete with models with parallax and cannot be excluded easily (e.g., see \cite{miyake2012}, \cite{koshimoto2017}, \cite{rota2021}, \cite{satoh2023}). { The Xallarap+Parallax model could be ruled out if the source has a mass incompatible with the observed flux in follow-up imaging \citep{bhattacharya2017, bhattacharya2020, blackman2021}. Furthermore, during light curve fitting, if a Xallarap+Parallax model returns an orbital period of 1 year, there is a high chance that the fit has converged to a mirror solution of a parallax model \citep{dong2009, hwang2011} and then the Xallarap+Parallax model can be discarded. But this is not the case with our model. Also, assuming that the lens is also a main-sequence star, we are finding the brightness of the lens consistent with the blend and source position on the CMD. So, it is difficult to rule out the Xallarap+Parallax model easily.} It also is to be noted that all the studies that were previously published using the 
Xallarap analysis to date are binary-lens events and this is the first study using a single-lens model.\\
\noindent In this study, we assume that the lens is single because the light curve resembles a Pa\'czynski curve. If a binary lens would have to produce a similar signal, then assuming the orbital period of $\sim$ 0.8 years and the mass of the lens as 0.1$M_{\odot}$, this would imply a separation (s) of 0.4 A.U. between the lens components. At such a separation, the line caustic formed by a lower mass companion would not go undetected in the light curve. Computing the magnification of a binary lens requires four additional parameters: $s$, $\phi$, the mass ratio of the lens $q$, and $\rho$. It also requires solving fifth-order complex polynomials for every pair of $s$ and $q$ in the parameter space. In addition, parameters $\dot{s}$ and $\dot{\phi}$ also enter the parameter list when the orbital motion of the lens effect is added to the modeling. Assuming a trajectory angle ($\phi$)such that the source goes away from the caustic structure would imply a single lens event. Then, involving a secondary lens becomes unnecessary. So, in this work, we have not modeled the lightcurves using a binary lens with an orbital motion effect.\\
We can expect many such events using the Nancy-Roman Space Telescope mission. While most of the focus of the microlensing community today is to analyze binary or tertiary events, single-lens events with asymmetric light curves need attention to find hidden signatures like OGLE-2017-BLG-0103. It is difficult to address any Xallarap solution as the source is faint, cannot be resolved using the current survey telescopes, and is usually in the bulge. So, a program for future AO observations using powerful large aperture telescopes like the Thirty Meter Telescope \citep{sanders2013, skidmore2015} can be made to study the degenerate solutions. Such observations can potentially validate the lens's mass and distance obtained using the galactic model constrained by various microlensing parameters.
\section*{Acknowledgement}
\noindent Sarang Shah is thankful to the India-TMT Coordination Center at the Indian Institute of Astrophysics for providing the time and resources to complete the analysis of this microlensing event. This work started as a part of the PhD thesis submitted to the University of Canterbury. This research has made use of data (\url{https://kmtnet.kasi.re.kr/ulens/}) from the KMTNet system (Kim et al. 2016) operated by the Korea Astronomy and Space Science Institute (KASI) at three host sites of CTIO in Chile, SAAO in South Africa, and SSO in Australia. Data transfer from the host site to KASI was supported by the Korea Research Environment Open NETwork (KREONET).
\newpage
\appendix
\section{Autocorrelation results of Xallarap and Parallax models}\label{appendix:autocorr}
In this section, we show the auto-correlation of the residuals calculated for the whole duration of the lightcurve for the KMTNet dataset. The residuals are the difference between the Xallarap+Parallax model and the observed data. The auto-correlation is 1 at lag 0, which is expected. However, as the lag value increases, the auto-correlation values for all the datasets drop and are consistently close to 0 and within the $95\%$ confidence interval throughout. 

\begin{figure}[htbp]
    \centering{
        \includegraphics[width=0.25\textwidth]{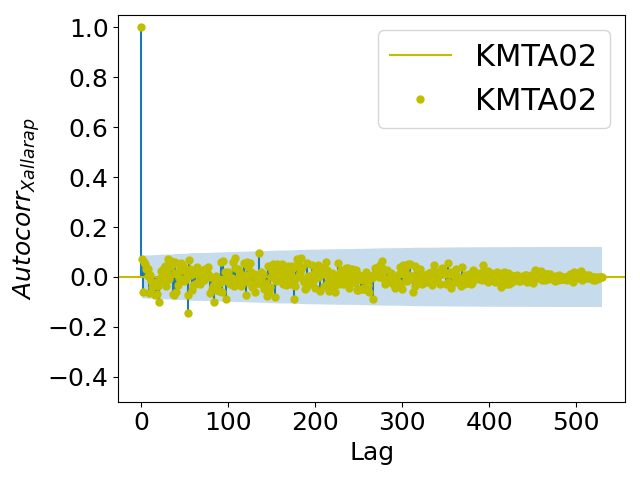}
    }
{
        \includegraphics[width=0.25\textwidth]{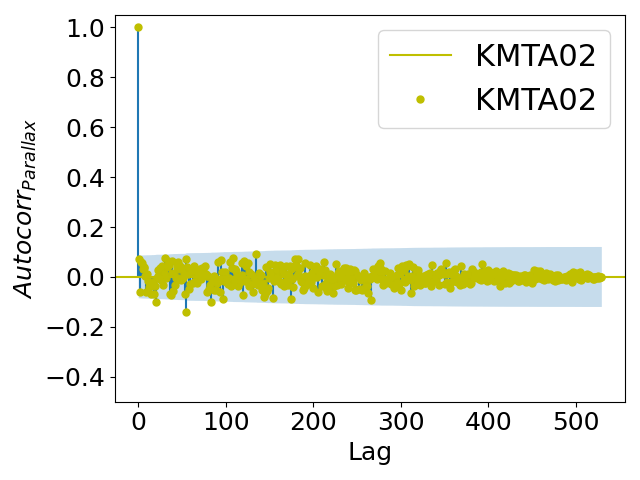}
    }\\
{
        \includegraphics[width=0.25\textwidth]{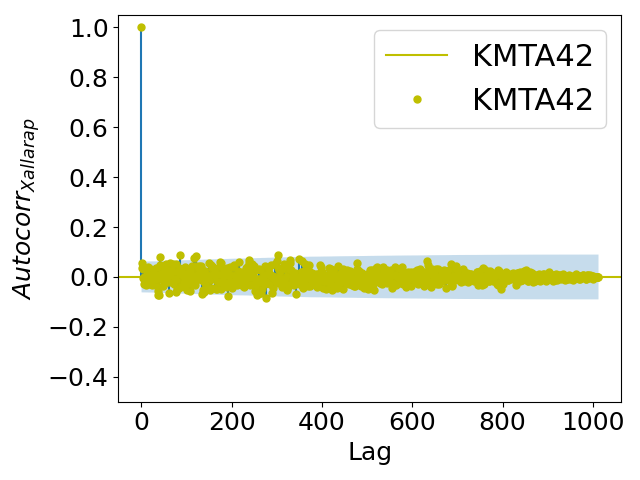}
    }
    {
        \includegraphics[width=0.25\textwidth]{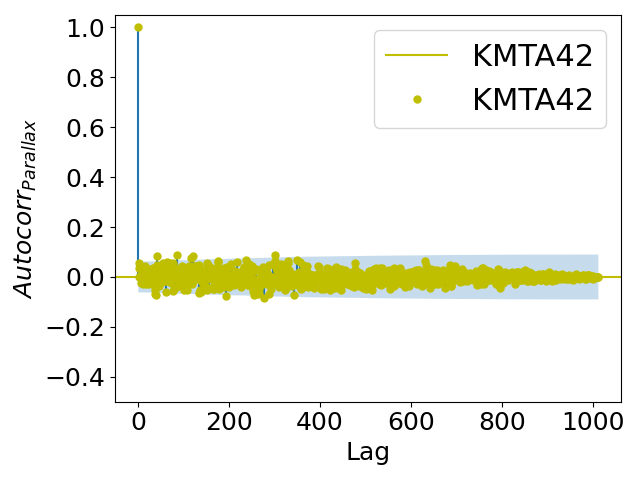}
    }\\
    {
        \includegraphics[width=0.25\textwidth]{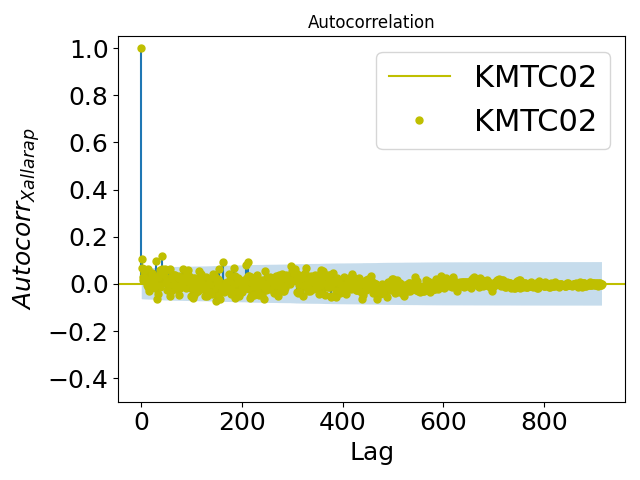}
    }
    {
        \includegraphics[width=0.25\textwidth]{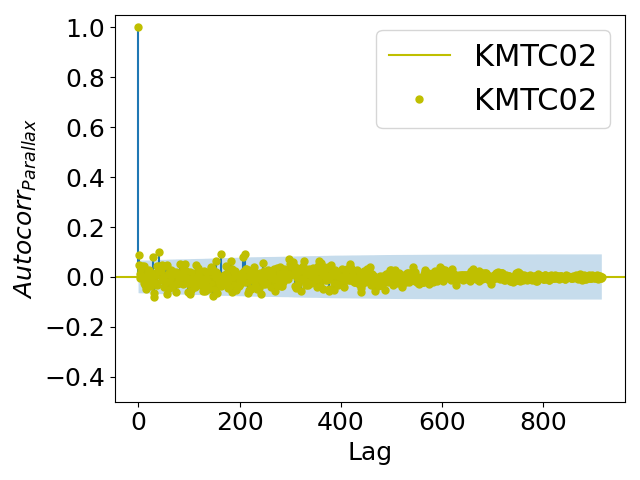}
    }\\
    {
        \includegraphics[width=0.25\textwidth]{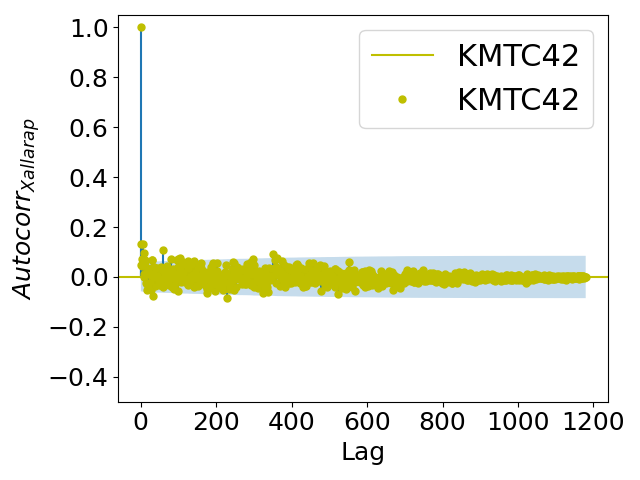}
    }
    {
        \includegraphics[width=0.25\textwidth]{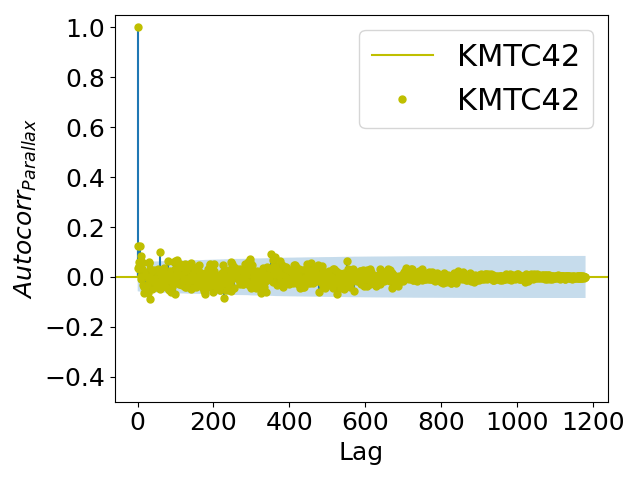}
    }\\
    {
        \includegraphics[width=0.25\textwidth]{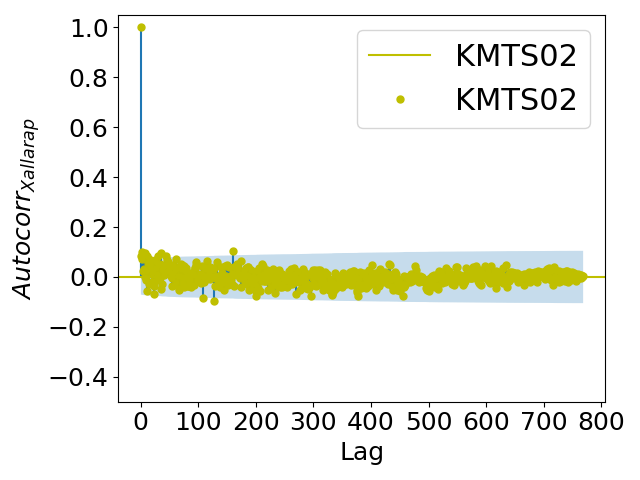}
    }
    {
        \includegraphics[width=0.25\textwidth]{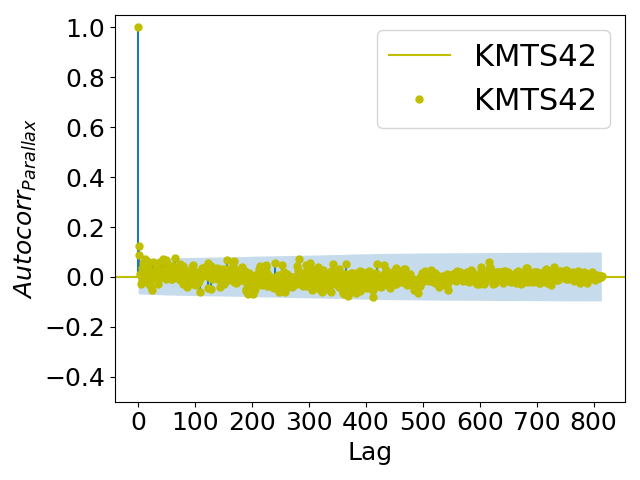}
    }
    \caption{Autocorrelation plots for different data sets. The autocorrelation is computed for the residuals of the best-fitted Xallarap+Parallax model for the lightcurve of OGLE-2017-BLG-0103.}
    \label{fig:autocorrelation_plots}
\end{figure}

\newpage
\bibliographystyle{aasjournal}
\bibliography{main}

\end{document}